\documentclass[twocolumn,aps,prd,showpacs,nofootinbib]{revtex4}
\usepackage{graphicx}
\usepackage[usenames]{color}
\usepackage{psfrag}
\usepackage[ps2pdf,colorlinks,bookmarks]{hyperref}

\definecolor{linkblue}{rgb}{0,0,0.8}
\definecolor{linkgreen}{rgb}{0,0.5,0}
\hypersetup{pdfpagemode=None, pdfstartview=FitH, linkcolor=linkblue, %
            citecolor=linkgreen, urlcolor=linkblue}
\def\beq{\begin{equation}}
\def\eeq{\end{equation}}
\def\bea{\setlength\arraycolsep{1.4pt}\begin{eqnarray}}
\def\eea{\end{eqnarray}}
\def\bit{\begin{itemize}}
\def\eit{\end{itemize}}

\def\ie{{i.e.}}
\def\eg{{e.g.}}

\def\ld{\left}
\def\rd{\right}
\def\ra{\rightarrow}

\def\fr{\frac}

\def\Del{\Delta}
\def\del{\delta}

\def\Lam{\Lambda}

\def\sig{\sigma}

\def\Om{\Omega}

\def\O{{\cal O}}

\def\hsf{hypersurface}
\def\hsfs{hypersurfaces}
\def\pert{perturbation}
\def\perts{perturbations}

\def\bra{\langle}
\def\ket{\rangle}

\def\Pdi{{\cal P_{\delta_{\rm i}}}}
\def\Pd{{\cal P_{\delta}}}

\def\camb{\textsc{camb}}

\def\omml{\Omega^{\rm loc}_{\rm m,0}}

\def\lfrw{$\Lambda$FLRW}

\def\rmd{d}

\begin{document}

\title[Linear kSZ effect and void models for acceleration]%
{Linear kinetic Sunyaev-Zel'dovich effect and void models for acceleration}

%\author{J. P. Zibin and A. Moss}
%\address{Department of Physics and Astronomy, %
%University of British Columbia, %
%Vancouver, BC, V6T 1Z1  Canada}
%\eads{\mailto{zibin@phas.ubc.ca}, \mailto{adammoss@phas.ubc.ca}}

\author{James P. Zibin} 
\email{zibin@phas.ubc.ca}
\affiliation{Department of Physics and Astronomy, %
University of British Columbia, %
Vancouver, BC, V6T 1Z1  Canada}

\author{Adam Moss} 
\email{adammoss@phas.ubc.ca}
\affiliation{Department of Physics and Astronomy, %
University of British Columbia, %
Vancouver, BC, V6T 1Z1  Canada}

\date{\today}

\begin{abstract}
   There has been considerable recent interest in cosmological 
models in which the current apparent acceleration is due to a very large 
local underdensity, or {\em void,} instead of some form of dark energy.  
Here we examine a new proposal to constrain such models using the linear 
kinetic Sunyaev-Zel'dovich (kSZ) effect due to structure within the void.  
The simplified ``Hubble bubble'' models previously studied appeared to 
predict far more kSZ power than is actually observed, independently of the 
details of the initial conditions and evolution of perturbations in such 
models.  We show that the constraining power of the kSZ effect is considerably 
weakened (though still impressive) under a fully relativistic treatment of 
the problem, and point out several theoretical ambiguities and observational 
shortcomings which further qualify the results.  Nevertheless, we conclude 
that a very large class of void models is ruled out by the combination of 
kSZ and other methods.
\end{abstract}

\pacs{98.80.Es, 95.36.+x, 98.65.Dx}
%\submitto{\CQG}

%\noindent\today

\maketitle

%The words table and figure should be written in full and not abbreviaged to 
%tab. and fig.  Do not include 'eq.' 'equation' etc before an equation number 
%or 'ref.' 'reference' etc before a reference number.

%IOP journals use a Roman d for a differential d, a Roman e for an exponential 
%e and a Roman i for the square root of -1. To accommodate this and to 
%simplify the typing of equations we have provided some extra definitions.
%\rmd, \rme and \rmi now give Roman d, e and i respectively

%Subscripts and superscripts should be in Roman type if they are labels rather 
%than variables

%Using the eqnarray environment equations will naturally be aligned left 
%without the use of any ampersands for alignment

%Where some secondary alignment is needed, for instance a second part of an
%equation on a second line, a single ampersand is added at the point of 
%alignment in each line

%Note that footnotes to the text should not be included in reference list, 
%but should appear at the bottom of the relevant page by using the \footnote 
%command.

%In addition to the standard \ref{<label>} the abbreviated forms given in the
%table 3 are available for reference to standard parts of the text

\section{Introduction}
\label{introsec}

   Probably the most surprizing cosmological discovery in recent decades 
has been the unexpected faintness of distant Type Ia supernovae 
(SNe) (see~\cite{ps03} for a review of the discovery).  This has been 
widely interpreted as evidence for the accelerated expansion of the Universe 
due to a cosmological constant (or similarly behaving ``dark energy''), 
or due to a modification to general relativity on very large scales.

   The standard picture of an accelerating, homogeneous and 
isotropic Friedmann-Lema\^itre-Robertson-Walker (FLRW) Universe is now 
supported by a wide variety of observations in addition to SNe 
(see~\cite{fth08,linder08} for reviews of the evidence).  Nevertheless, 
uncertainty about the fundamental origin of a small cosmological 
constant has led to considerable effort in studying alternative models 
in which inhomogeneity plays a key role.  The 
simplest of these, both conceptually and practically, is the idea 
that an observer in an approximately 
spherical underdensity, or {\em void,} which extends to redshift 
$z \simeq 1$, will observe an {\em apparent} acceleration due to 
the inhomogeneity~\cite{tomita00,gtbgo99,celerier99}.  These 
models dispense with the need for a mysterious dark energy component 
at the cost of a violation of the Copernican principle, since the 
observer must be very close to the centre of the void in order 
to satisfy cosmic microwave background (CMB) dipole 
constraints~\cite{moffat95,nakao95,tomita96,hmm97}.  Although 
void models (and others based on inhomogeneity) potentially 
solve the coincidence problem with the standard \lfrw\ model\footnote{We 
refer to the standard cosmological model as \lfrw\ to emphasize its geometry, 
whereas the more common term ``$\Lam$ cold dark matter'' ($\Lam$CDM) 
emphasizes its matter content.} by linking 
the appearance of acceleration to the formation of nonlinear structures, 
they are not free of temporal tuning~\cite{fmzs10}.

   Considerable effort has gone into confronting such void models for 
acceleration with a variety of different data (see, \eg, the 
reviews~\cite{mn11,bck11} in this focus section).  SN data can always 
be satisfied by choosing the radial profile of a void to match the 
redshift-luminosity distance relation of \lfrw~\cite{ykn08}; 
nevertheless, these data do put crucial constraints on the depth and 
width of a void.  Another class of observations involves the properties 
of structures within the void, \eg\ studies of the baryon acoustic 
oscillations (BAO)~\cite{gbh08,zms08}.  While potentially very constraining, 
these approaches are hampered by the difficulty of predicting the evolution 
of perturbations on LTB backgrounds~\cite{z08,ccf09}, although in special 
cases, such as near the centre, the evolution can be followed analytically 
and strong constraints made~\cite{mzs10}.  More fundamentally, these 
approaches are hindered by our ignorance of the initial conditions (ICs) 
for the perturbations within the void: since the origin of the void 
itself is unclear, we cannot be sure that the ICs which apply at the 
radius of last scattering (and are manifest in the CMB) also apply locally.

   The primary anisotropies in the CMB contain a wealth of information, 
and hence their ability to constrain void models has been carefully 
studied.  As first pointed out in~\cite{zms08}, as long as the primordial 
spectrum is close to scale invariant, observations of the CMB temperature 
anisotropy spectrum can only be satisfied in void models if the {\em local} 
expansion rate is so low as to rule the models out.  These results were 
shown to persist when the void is embedded in a spatially curved 
background~\cite{bnv10,mzs10,mp10}.

   The CMB is also expected to contain a variety of secondary anisotropies 
generated at late times.  Galaxy clusters within the void will see 
very large CMB dipoles and hence are expected to produce substantial 
anisotropies via the kinetic Sunyaev-Zel'dovich (kSZ) effect~\cite{sz80}, 
as first suggested in~\cite{vfw06}.  This has been used to put constraints 
on void models in~\cite{gbh08b,yns10b}.  Recently, a related approach 
in which the kSZ anisotropies due to all structure within the void 
are considered was introduced by Zhang and Stebbins~\cite{zs10} 
(hereafter ZS10).  The authors found that void models predicted far 
more kSZ power than is actually observed at small angular scales, 
and hence claimed that all such models were ruled out.  However, ZS10 
employed a simplified non-relativistic void model and did not 
examine the dependence of the effect on a variety of parameters.

   In this work we aim to repeat the analysis of ZS10, but in a fully 
consistent general relativistic framework.  We employ the exact, 
spherically symmetric pressureless matter solution to Einstein's 
equations, known as the Lema\^itre-Tolman-Bondi (LTB) 
spacetime~\cite{lemaitre33,tolman34,bondi47}.  We assume that the LTB 
model contains no decaying mode, since we can then maintain the 
standard picture of an early homogeneous Universe consistent with 
inflation.  We also ignore the possibility of a large, early isocurvature 
mode between radiation and matter.  We assume, with ZS10, 
that the matter power spectrum today is that of the standard \lfrw\ 
model, hence bypassing all uncertainty related to the ICs and evolution 
of perturbations inside the void.  However, we show that the 
question of which {\em scales} the spectrum is to be evaluated at 
is not trivial.  We confirm that void models which satisfy the SN data 
overpredict kSZ power, though considerably less so than indicated in ZS10, 
and point out that there are numerous theoretical ambiguities and 
observational uncertainties that affect the reliablity of kSZ calculations 
in void models.

   We begin in section~\ref{kSZcalcsec} with a calculation of the kSZ 
power in LTB models.  We point out that there is considerable ambiguity 
in relating length scales in LTB and standard \lfrw\ models, but 
propose a resolution.  Next, in section~\ref{kSZbehavsec}, we illustrate 
kSZ spectra and the distribution of power in wave number and redshift space.  
Section~\ref{constraintsec} presents void model constraints using SN and 
local Hubble rate data.  Section~\ref{robustsec} examines the dependence of 
the kSZ power on various parameters.  We extend our results to a calculation 
of the Compton $y$-distortion in section~\ref{imptltbsec}.  
Our conclusions are presented in 
section~\ref{conclsec}.  The \hyperref[LTBsec]{Appendix} presents a brief 
description of our LTB models.  Throughout this paper we set $c = 1$ 
and define the conventional dimensionless Hubble rate $h_0$ via $H_0 \equiv 
100\, h_0 \, {\rm km} \, {\rm s^{-1}} \, {\rm Mpc^{-1}}$.

\section{Calculating the linear kinetic Sunyaev-Zel'dovich effect}
\label{kSZcalcsec}
\subsection{Derivation of the kSZ power}
\label{derkSZsec}

   The kSZ effect~\cite{sz80} due to Thomson scattering of the CMB from free 
electrons is conventionally written as
\beq
\fr{\Del T(\mathbf{n})}{T} = \int\mathbf{v}(\mathbf{n},z)\cdot\mathbf{n}\,
   [1 + \del(\mathbf{n},z)]\rmd\tau,
%\label{kSZdeffull}
\eeq
where $\mathbf{n}$ is the line-of-sight direction, $\del(\mathbf{n},z) 
\equiv \del\rho_e(\mathbf{n},z)/\rho_e(z)$ is the comoving free electron 
density perturbation in direction $\mathbf{n}$ and at redshift $z$, and 
$\tau$ is the optical depth along the line of sight.  The quantity 
$\mathbf{v}(\mathbf{n},z)$ is the relative velocity between the free 
electrons and the CMB rest frame (the frame in which the CMB dipole 
vanishes) at $(\mathbf{n},z)$.  A scatterer at 
$(\mathbf{n},z)$ will observe a {\em radial} component of a CMB dipole 
directly related to the radial component $\beta(\mathbf{n},z) \equiv 
\mathbf{v}(\mathbf{n},z)\cdot\mathbf{n}$.\footnote{In terms of the 
spherical harmonic coefficients $a_{\ell m}$ to be defined in (\ref{almdef}), 
we have $\beta(\mathbf{n},z) = \sqrt{3/(4\pi)}a_{10}(\mathbf{n},z)$, 
when the polar axis is aligned along $\mathbf{n}$.}

   In an FLRW background, the dipole $\beta(\mathbf{n},z)$ vanishes by 
isotropy.  Therefore in realistic models with structure, $\beta(\mathbf{n},z)$ 
is a perturbative quantity, corresponding to peculiar velocities.  In an 
inhomogeneous LTB background, on the other hand, the lack of isotropy away 
from the centre means that the dipole will not generally vanish.  Therefore, 
in a model with structure on top of an LTB background, we can decompose 
the dipole according to $\beta(\mathbf{n},z) = \bar\beta(z) + 
\del\beta(\mathbf{n},z)$, with the first part due to the LTB background (a 
very-large-scale ``bulk velocity''), and the second part due to the 
superimposed perturbative structure (peculiar velocities relative to the 
LTB background).  Therefore we can write the kSZ anisotropy as
\beq
\fr{\Del T(\mathbf{n})}{T} = \int[\beta(z) + \del\beta(\mathbf{n},z)]
                        [1 + \del(\mathbf{n},z)]\rmd\tau
\label{kSZdeffull}
\eeq
(we henceforth drop the bar over $\beta(z)$).  We initially assume that the 
free electron fluctuations match those of the 
total matter, so that $\del\rho_e/\rho_e = \del\rho_{\rm m}/\rho_{\rm m}$, 
but consider the validity of this approximation in section~\ref{baryonsec}.  
The dipole and density \perts\ are evaluated on the light cone, so, \eg,
\beq
\del(\mathbf{n},z) = \del(\mathbf{n},t(z),r(z))
\eeq
for time coordinate $t$ and radial coordinate $r$.  We will often use the 
shorthand notation
\beq
Q(r) = Q(t(z),r(z)) = Q(z)
\eeq
or
\beq
Q(t) = Q(t(z),r(z))
\eeq
for quantities $Q(t,r)$ evaluated on the observer's past light cone.

   As we mentioned above, in FLRW backgrounds the background dipole 
vanishes, $\beta(z) = 0$.  In this case, the linear term in 
(\ref{kSZdeffull}), $\int\del\beta\rmd\tau$, is negligible due to a 
geometrical cancellation~\cite{kaiser84}.  Therefore the second order term 
$\int\del\beta\del\rmd\tau$ dominates, and the effect in that case is 
sometimes known as the Ostriker-Vishniac effect~\cite{ov86}.  In LTB 
void models, the term $\int\beta\rmd\tau$ contributes only a small 
monopole\footnote{However, this term should lead to higher multipole 
anisotropies in nonspherical voids.} and hence we will ignore it.  We 
will assume that the term $\int\del\beta\rmd\tau$ vanishes (or at least 
that it is subdominant) in LTB models as it does in the FLRW case.  
Hence we are left with one relevant first order term,
\beq
\fr{\Del T(\mathbf{n})}{T} = \int\beta(z)\del(\mathbf{n},z)\rmd\tau.
\label{kSZdef}
\eeq
The fact that~(\ref{kSZdef}) is {\em linear} in the fluctuation amplitude 
leads us to call it the {\em linear} kSZ effect, after ZS10, and will result 
in much larger kSZ power in void models than in FLRW models.  It is 
important to stress that the linear kSZ effect is fundamentally different 
from the nonlinear effect in FLRW models.  Because the linear kSZ effect 
does not contain dipole perturbations (\ie\ peculiar velocities), it is 
independent of the details of the velocity power spectrum.  On the other 
hand, the velocity power is crucial in the calculation of kSZ in FLRW 
models.  The insensitivity to the velocity power removes a significant 
source of uncertainty in the LTB kSZ calculation, although considerable 
ambiguities will remain, as we will see.

   In order to evaluate the kSZ integral~(\ref{kSZdef}), we will wish to 
perform a harmonic decomposition of the \pert\ field $\del(\mathbf{n},z)$.  
To be able to do this, the field should be defined on a flat (or at least 
constant-curvature) spacelike \hsf.  However, flat slices are not natural 
in general LTB spacetimes: they will not generally be orthogonal to the 
comoving worldlines, so \hsfs\ of constant proper time will not be of 
constant curvature.  Nevertheless, recalling that we are 
restricting the LTB models to growing mode profiles, we can bypass this 
difficulty by relating the \pert\ field on the past light cone, 
$\del(\mathbf{n},z)$, to that on some slice $t_{\rm i}$, early enough that the 
LTB spacetime is close to FLRW at that time.  We can then perform a 
decomposition on the early slice.  Therefore, we introduce a growth 
function $D(t)$ to relate the density \perts\ at $t_{\rm i}$, 
$\del_{\rm i}(\mathbf{n},r) \equiv \del(\mathbf{n},t_{\rm i},r)$, to those at 
some later time via
\beq
\del(\mathbf{n},t(z),r(z)) = D(t(z))\del_{\rm i}(\mathbf{n},r(z)).
\label{growthdef}
\eeq
By virtue of this expression, the coordinate $r$ must be comoving.  For 
the case of linear fluctuations about an FLRW background dominated by 
dust and cosmological constant at late times, we have the simple relation
\beq
D(t) = \fr{a(t)}{a(t_{\rm i})}\fr{g(t)}{g(t_{\rm i})},
\eeq
where $a(t)$ is the FLRW scale factor and $g(t)$ the usual growth 
suppression factor.

   It is very important to point out that the relation~(\ref{growthdef}) 
neglects a couple of significant physical effects.  First, in relating the 
perturbation at some coordinates $(\mathbf{n},t,r)$ to that on the same 
comoving worldline at $(\mathbf{n},t_{\rm i},r)$, it ignores any scale 
dependence in the evolution.  In particular, at late times the nonlinear 
growth is expected to be greatest on the smallest scales.  The second effect 
is that the scalar perturbation evolution on the LTB background is expected 
to couple to vector and tensor modes, necessarily coupling different 
comoving worldlines.  However, in the approximation that the coupling 
to tensors is ignored, it can be shown that the evolution satisfies a 
relation similar to~(\ref{growthdef})~\cite{z08}.  More importantly, as 
mentioned in the \hyperref[introsec]{Introduction}, we will not attempt a 
description of 
the perturbation ICs and evolution in this work; instead, we will 
assume that the actual matter power spectrum on the light cone is close 
to that of the standard \lfrw\ model.  Therefore, we will consider it 
reasonable to ignore this latter effect.  On the other hand, it will 
be necessary to deal with the former effect in an ad hoc manner by 
introducing scale dependence below.

   The LTB growing mode assumption allows us to expand the field 
$\del_{\rm i}(\mathbf{n},r)$ in spherical harmonic functions, 
$Y_{\ell m}(\mathbf{n})$, and spherical Bessel functions of the first kind, 
$j_\ell(kr)$, according to
\beq
\del_{\rm i}(\mathbf{n},r) = \sqrt{\fr{2}{\pi}}\int\rmd k\,k
    \sum_{\ell m}\del_{{\rm i},\ell m}(k)j_\ell(kr)Y_{\ell m}(\mathbf{n}),
\label{sphexpn}
\eeq
where $k$ is the wave number.  It is important to stress the meaning of 
the quantities $r$ and $k$ here: the radial comoving coordinate $r$ must 
be proportional to {\em proper} distance at $t_{\rm i}$ (at least at 
background level, on the nearly FRLW background), and the wave number 
$k$ is similarly proportional to proper wave number, in order that the 
harmonic decomposition be valid.

\begin{widetext}
   Now, substituting~(\ref{sphexpn}) and~(\ref{growthdef}), we can 
write~(\ref{kSZdef}) as
\beq
\fr{\Del T(\mathbf{n})}{T} = \sqrt{\fr{2}{\pi}}\int\rmd rF(r)D(t(r))\int\rmd k
    \,k\sum_{\ell m}\del_{{\rm i},\ell m}(k)j_\ell(kr)Y_{\ell m}(\mathbf{n}),
\label{DelTkSZsi}
\eeq
where
\beq
F(r) \equiv \beta(r)\fr{\rmd\tau}{\rmd r}.
\eeq
As we mentioned above, this expression for the kSZ anisotropy ignores 
scale dependence in the perturbation evolution.  However, as we will see 
below, the matter power on nonlinear scales will be crucial to the 
strength of the kSZ technique in constraining LTB models.  Therefore, 
in order to capture the effect of the nonlinear power, we must promote 
the growth function to be scale dependent, $D(t) \ra D(t,k)$.  With this 
replacement, (\ref{DelTkSZsi}) becomes
\beq
\fr{\Del T(\mathbf{n})}{T}
   = \sqrt{\fr{2}{\pi}}\int\rmd rF(r)\int\rmd k\,kD(t(r),k)
     \sum_{\ell m}\del_{{\rm i},\ell m}(k)j_\ell(kr)Y_{\ell m}(\mathbf{n}).
\eeq
\end{widetext}
To avoid this ad hoc procedure would require developing the theory of 
perturbations on LTB backgrounds.  While some progress has been made in 
the linear case~\cite{z08,ccf09}, the nonlinear case has not yet been 
addressed.  However, although strictly the growth function $D(t,k)$ is 
ill-defined since we cannot perform a harmonic decomposition at late 
times, this procedure might be justified in that it may 
be possible to consider such a decomposition valid on the very 
small scales (much smaller than the LTB curvature scale) relevant to 
the kSZ effect.  This leads to our first major caveat: our result will 
likely be invalid on the largest angular scales, and the degree of 
accuracy will be uncertain even on small scales.

   Next we can calculate the spherical harmonic coefficients of the observed 
temperature field,
%\bea
%a_{\ell m} &\equiv\int\fr{\Del T(\mathbf{n})}{T}Y_{\ell m}^*(\mathbf{n})\rmd\Om
%\label{almdef}\\
%               &= \sqrt{\fr{2}{\pi}}\int\rmd rF(r)
%                  \int\rmd k\,kD(r,k)\del_{{\rm i},\ell m}(k)j_\ell(kr).
%\eea
\bea
\hspace{-0.5cm}
a_{\ell m}&\equiv&\int\fr{\Del T(\mathbf{n})}{T}Y_{\ell m}^*(\mathbf{n})\rmd\Om
\label{almdef}\\
                &=& \sqrt{\fr{2}{\pi}}\int\rmd rF(r)
                   \int\rmd k\,kD(r,k)\del_{{\rm i},\ell m}(k)j_\ell(kr).
\eea
Finally, the kSZ power at multipole $\ell$ is given by
%\beq
%C_\ell \equiv \bra a_{\ell m}a_{\ell m}^*\ket
%       = 4\pi\int\fr{\rmd k}{k}\Pdi(k)\ld(\int\rmd rF(r)D(r,k)j_\ell(kr)\rd)^2.
%\label{Clexact}
%\eeq
\bea
\hspace{-0.5cm}
C_\ell &\equiv& \bra a_{\ell m}a_{\ell m}^*\ket\\
     &=& 4\pi\int\fr{\rmd k}{k}\Pdi(k)\ld(\int\rmd rF(r)D(r,k)j_\ell(kr)\rd)^2.
\label{Clexact}
\eea
In deriving this final expression, we have used the defining relation
\beq
\bra \del_{{\rm i},\ell m}(k)\del_{{\rm i},\ell'm'}^*(k')\ket =
\fr{2\pi^2}{k^3}\Pdi(k)\del(k - k')\del_{\ell\ell'}\del_{mm'}
\eeq
for the power spectrum $\Pdi(k)$ of Gaussian random matter field 
$\del_{\rm i}(\mathbf{n},r)$.  The spectrum $\Pdi(k)$ is well defined since 
a harmonic decomposition is possible at $t_{\rm i}$, when the spacetime 
is near FLRW.

   Equation~(\ref{Clexact}) provides a general expression for the kSZ 
power on LTB backgrounds, subject to the major caveat regarding the 
ambiguity of defining the growth factor $D(r,k)$.  The factor $F(r)$ in the 
integrand will depend on the particular radial LTB profile chosen.  It 
can be written
\beq
F(r) = \beta(r)\fr{\rmd\tau}{\rmd z}\fr{\rmd z}{\rmd r},
\label{Fdef}
\eeq
where
\beq
\fr{\rmd\tau}{\rmd z}
   = \fr{\sig_{\rm T}f_{\rm b}(2 - Y_{\rm He})\rho_{\rm m}(z)}
        {2m_p(1 + z)H_\parallel(z)}
\label{dtaudz}
\eeq
and
\beq
\fr{\rmd z}{\rmd r} = (1 + z)H_\parallel(z)\fr{Y'}{\sqrt{1 - K}}.
\label{dzdr}
\eeq
Here $\sig_{\rm T}$ is the Thomson cross section, $f_{\rm b} \equiv 
\rho_{\rm b}/\rho_{\rm m}$ is the baryon fraction, $Y_{\rm He}$ is the 
helium mass fraction, $m_p$ is the proton mass, and the remaining LTB 
functions are defined in the \hyperref[LTBsec]{Appendix}.  This 
expression applies 
after reionization, when we have assumed that the baryonic matter is 
completely ionized.  We use the value $Y_{\rm He} = 0.24$ throughout 
this work, and nominally set $f_{\rm b} = 0.168$, as implied by CMB 
observations~\cite{wmap7params}.  As we discuss in detail in 
section~\ref{varyparamsec}, it is likely that this standard value of 
baryon fraction should be modified locally in void models, leading to 
considerable uncertainty in our kSZ predictions.

\subsection{The matter power spectrum}
\label{matterPSsec}

   After the LTB profile is specified, the remaining quantities required 
to calculate the kSZ power are the 
matter power spectrum $\Pdi(k)$ and the growth function $D(r,k)$.  
As explained above, we do not attempt to calculate these quantities from 
ICs; rather, we begin by assuming that the actual matter power along our 
past light cone matches that of the standard \lfrw\ model.  Later we will 
consider the effect of relaxing this assumption.  However, implementing 
even this simple prescription for the matter power is far from trivial.  
First, as already mentioned in section~\ref{derkSZsec}, it is not clear 
how to define a power spectrum at late times when a harmonic decomposition 
is not possible.  But there is a second important reason: the wave numbers 
$k$ with respect to which the power spectra are specified have different 
meanings and behaviours in LTB and FLRW models.  Since the matter power 
spectrum can depend sensitively on scale, this will lead to a further 
significant ambiguity in specifying the matter power.

   To see this, note first of all that, for FLRW models, the power 
spectrum on the past 
light cone, $\Pd(k,z)$, is usually specified in terms of {\em comoving} 
wave number $k$ (or $k/h_0$), which is equivalent to {\em proper} wave 
number today.  In general, perturbation modes in LTB and FLRW models 
with identical proper wave numbers today will not share identical proper 
wave numbers on the past light cone, due to the very different background 
evolution of the two models.  (Recall that the LTB and standard \lfrw\ 
background evolutions differ at {\em zeroth} order.)  Therefore it is 
certainly incorrect to state that the LTB model should share the same 
$\Pd(k,z)$ at the same proper wave number {\em today} as the standard 
\lfrw\ model.

   However, it is {\em also} generally incorrect to require that the power 
$\Pd(k,z)$ be the same in both models at the same proper $k$ specified 
at the {\em same redshift} $z$ on the past light cone.  The reason is that 
what is actually observed in galaxy surveys is not proper wave numbers at 
some $z$, but rather angular and redshift separations.  A pair of 
galaxies with the same proper separation at the same $z$ in LTB and FLRW 
models will generally be observed with very different angular and 
redshift separations in the two models, again due to the different 
background geometry.

   Therefore, what we require is that the LTB model has the same matter 
power on the same angular and redshift scales, at the same $z$, as the 
standard \lfrw\ model.  In any model, for a mode oriented perpendicular to 
the line of sight with {\em proper} wave number $k_\perp$ at redshift $z$, 
the corresponding angular scale on the sky is
\beq
\Del\theta = \fr{\alpha}{k_\perp(z)d_{\rm A}(z)},
\label{Deltatheta}
\eeq
where $d_{\rm A}(z)$ is the angular diameter distance to $z$, measured as a 
{\em proper} distance at $z$, and $\alpha$ is a constant (independent 
of the model) of order unity.  Therefore, if the two models are to share 
the same power at the same angular scales, they must have the same power 
at proper wave numbers related by
\beq
k_\perp^{\rm LTB}(z) = k_\perp^\Lam(z)
                      \fr{d_{\rm A}^\Lam(z)}{d_{\rm A}^{\rm LTB}(z)}.
\label{kktheta}
\eeq
In practice, if the LTB model is to fit the Type Ia supernova data and 
measurements of the local Hubble rate, then $d_{\rm A}^{\rm LTB}(z)$ must be 
similar to $d_{\rm A}^\Lam(z)$, at least out to $z \simeq 1$, and hence a 
first approximation would be to set $k^{\rm LTB}(z) = k^\Lam(z)$.

   Similarly, if the two models are to share the same power at the same 
redshift scales, they must have the same power at proper wave numbers 
related by
\beq
k_\parallel^{\rm LTB}(z) = k_\parallel^\Lam(z)
                          \fr{H_\parallel^{\rm LTB}(z)}{H^\Lam(z)}.
\label{kkz}
\eeq
This leads directly to a fundamental problem: generically, we will have 
$k_\perp^{\rm LTB}(z) \ne k_\parallel^{\rm LTB}(z)$, when $k_\perp^\Lam(z) 
= k_\parallel^\Lam(z)$, due to the background geometry of the LTB 
models.\footnote{This same model dependence in the meaning of $\Pd(k,z)$ 
will also be present when comparing different {\em FLRW} models, although to 
a lesser degree than with LTB models.}  This means that there is no unique 
scale in \lfrw\ corresponding to a particular scale in the LTB model, and 
vice versa.  (Indeed, this discrepancy between 
angular and redshift scales in the two models results in the strength of 
the radial BAO scale in constraining LTB models~\cite{zms08,mzs10}.)

   To patch over this problem, we define an ``isotropized'' version of 
the relation between wave numbers in the two models by
%\bea
%k^\Lam(z) &\equiv \ld[\ld(k_\perp^\Lam(z)\rd)^2k_\parallel^\Lam(z)\rd]^{1/3}\\
%   &= \ld[\ld(
%    k_\perp^{\rm LTB}(z)\fr{d_{\rm A}^{\rm LTB}(z)}{d_{\rm A}^\Lam(z)}\rd)^2
%    k_\parallel^{\rm LTB}(z)\fr{H^\Lam(z)}{H_\parallel^{\rm LTB}(z)}\rd]^{1/3}.
%\label{kLTBLam}
%\eea
\bea
k^\Lam(z) &\equiv& \ld[\ld(k_\perp^\Lam(z)\rd)^2k_\parallel^\Lam(z)\rd]^{1/3}\\
   &=& \ld[\ld(
    k_\perp^{\rm LTB}(z)\fr{d_{\rm A}^{\rm LTB}(z)}{d_{\rm A}^\Lam(z)}\rd)^2
    k_\parallel^{\rm LTB}(z)\fr{H^\Lam(z)}{H_\parallel^{\rm LTB}(z)}\rd]^{1/3}
\label{kLTBLam}
\eea
This relation weights the transverse relation more heavily than the radial 
one, since there are twice as many transverse dimensions.  It is important 
to stress that the particular form of this relation we have chosen is 
still largely arbitrary.  [This notion of an 
isotropized scale is akin to the isotropized distance measure often used 
in the literature on BAO observations (\eg~\cite{eisenstein05})].  Our final 
statement, then, is that the LTB and \lfrw\ models must share the same 
matter power on scales related by~(\ref{kLTBLam}).  The ambiguity in 
this relation leads to our second major caveat: uncertainty in 
the $k$ scales, together with the sensitivity of the matter power to $k$ 
in standard spectra, will lead to further uncertainty in the calculated kSZ 
power.

   The evolution of proper wave numbers in the radial and transverse 
directions in the LTB model is straightforward to specify.  For a mode 
of proper wave number $k(t_{\rm i})$ specified at the time $t_{\rm i}$ 
early enough that the spacetime is near FLRW (and hence the power 
spectrum is presumably isotropic), we have
%\beq
%k_\perp(t,r) = k(t_{\rm i},r)\fr{Y(t_{\rm i},r)}{Y(t,r)}, \qquad
%k_\parallel(t,r) = k(t_{\rm i},r)\fr{Y'(t_{\rm i},r)}{Y'(t,r)}.
%\label{kevolnLTB}
%\eeq
\beq
k_\perp(t,r) = k(t_{\rm i},r)\fr{Y(t_{\rm i},r)}{Y(t,r)}, \quad
k_\parallel(t,r) = k(t_{\rm i},r)\fr{Y'(t_{\rm i},r)}{Y'(t,r)}.
\label{kevolnLTB}
\eeq
We stress again that our inability to perform proper harmonic 
decompositions of fields at late times implies that we must interpret 
relations of this sort loosely.

   Recall that the general expression for kSZ power~(\ref{Clexact}) 
involves an integral over {\em proper} $k$ at $t_{\rm i}$.  To determine 
the matter power on the corresponding scale, we first use~(\ref{kevolnLTB}) 
to relate the $k$ values at $t_{\rm i}$ to those on the light cone in the 
LTB model.  We then use~(\ref{kLTBLam}) to calculate the corresponding 
scale in the standard \lfrw\ model, using the fitting function of \cite{wu10} 
and the WMAP7 best-fit parameters~\cite{wmap7params} to evaluate 
$d_{\rm A}^\Lam(z)$ and $H^\Lam(z)$.  The \lfrw\ power can be calculated 
using any of several public codes; we used \camb~\cite{camb}, which optionally 
includes nonlinear power with a Halofit calculation~\cite{halofit}.  
Our matter spectrum used the normalization provided by CMB 
observations~\cite{wmap7params}, $\sig_8 = 0.81$, where $\sig_8$ is the 
linear amplitude on a scale of $8/h_0$ Mpc.  As we discuss in 
section~\ref{varyparamsec}, however, there is considerable ambiguity over 
this choice in the context of LTB models.  Finally, note that~(\ref{kLTBLam}) 
becomes ill-defined when $H_\parallel^{\rm LTB}(z) \le 0$.  This does 
in fact happen for the very deepest voids, leading to multi-valued 
distance-redshift relations~\cite{mbhe98}.  Therefore our kSZ calculations 
will not be valid in the multi-valued regime.

\subsection{The Limber approximation}

   We can significantly simplify the kSZ power expression~(\ref{Clexact}) 
using the so-called Limber approximation~\cite{limber53}.  This approximation 
is purely mathematical, as opposed to the physical approximations 
discussed previously.  To employ it, we substitute
\beq
j_\ell(kr) \simeq \sqrt{\fr{\pi}{2\ell + 1}}\del(\ell + 1/2 - kr)
\eeq
in~(\ref{Clexact}).  This substitution will be accurate to $\O(1/\ell^2)$ 
when the integrand $F(r)D(r,k)$ is slowly varying on the scale of the 
Bessel oscillations~\cite{als04}.  The result for the kSZ power is
%\bea
%C_\ell &\simeq \fr{4\pi^2}{2\ell + 1}
%               \int\fr{\rmd k}{k}F^2\ld(\fr{2\ell + 1}{2k}\rd)
%                    D^2\ld(\fr{2\ell + 1}{2k},k\rd)\fr{\Pdi(k)}{k^2}\\
%            &= \fr{16\pi^2}{(2\ell + 1)^3}
%               \int\rmd r\,rF^2(r)D^2\ld(r,\fr{2\ell + 1}{2r}\rd)
%                   \Pdi\ld(\fr{2\ell + 1}{2r}\rd).\label{Limberri}
%\eea
\newpage
\begin{widetext}
\bea
C_\ell &\simeq& \fr{4\pi^2}{2\ell + 1}
               \int\fr{\rmd k}{k}F^2\ld(\fr{2\ell + 1}{2k}\rd)
                    D^2\ld(\fr{2\ell + 1}{2k},k\rd)\fr{\Pdi(k)}{k^2}\\
            &=& \fr{16\pi^2}{(2\ell + 1)^3}
               \int\rmd r\,rF^2(r)D^2\ld(r,\fr{2\ell + 1}{2r}\rd)
                   \Pdi\ld(\fr{2\ell + 1}{2r}\rd).\label{Limberri}
\eea
\end{widetext}
If we attempt to define a matter power spectrum on the light cone by 
$\Pd(k,z) = D^2(z,k)\Pdi(k)$, then~(\ref{Limberri}) becomes
\beq
C_\ell \simeq \fr{16\pi^2}{(2\ell + 1)^3}
          \int\rmd r\,rF^2(r)\Pd\ld(\fr{2\ell + 1}{2r},z(r)\rd).
\label{Limberr}
\eeq
Note that all $k$ values in these expressions are proper wave numbers 
evaluated at $t_{\rm i}$.  To specify the standard \lfrw\ matter spectrum, 
we must relate these early time $k$ values to corresponding scales in 
the \lfrw\ model using the procedure described in section~\ref{matterPSsec}.

   Equation~(\ref{Limberr}) apparently has a simple interpretation as an 
integration down the past light cone of a weighted local matter power 
spectrum.  However, it is important to stress that the interpretation of 
the late time spectrum $\Pd(k,z)$ is highly ambiguous, as discussed in 
sections~\ref{derkSZsec} and~\ref{matterPSsec}.  The same replacement of 
$\Pdi(k)$ with $\Pd(k,z)$ cannot be made with the exact expression for 
kSZ power, (\ref{Clexact}), and hence neither can such a simple 
interpretation be made.  Note that~(\ref{Limberr}) is equivalent to the 
corresponding expression in ZS10, up to $\O(1/\ell)$.  The specific 
$\ell$-dependent prefactors in~(\ref{Limberr}) ensure that it is accurate 
to $\O(1/\ell^2)$.

\begin{figure}\begin{center}
%[ht]
%\includegraphics[width=0.7\columnwidth]{exact_Limber.eps}
\includegraphics[width=\columnwidth]{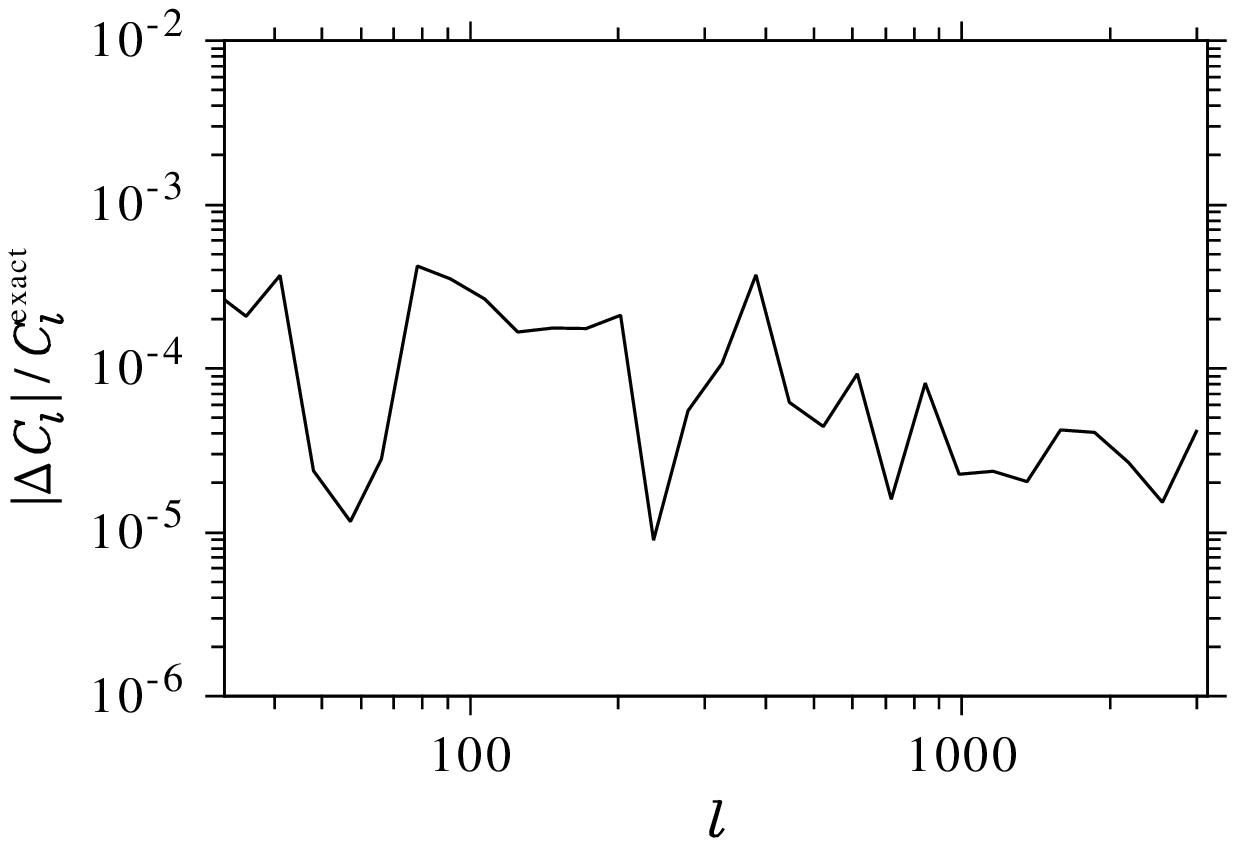}
\caption{The relative error of the Limber approximation given 
by~(\ref{Limberri}), $C_\ell^{\rm Limber}$, with respect to the exact kSZ 
power given by~(\ref{Clexact}), $C_\ell^{\rm exact}$.  $\Del C_\ell 
\equiv C_\ell^{\rm Limber} - C_\ell^{\rm exact}$.  Results are presented 
for the fiducial LTB model.
\label{exact_Limberfig}}
\end{center}\end{figure}

   The Limber approximation provides a significant advantage to numerical 
computations, especially if a large region of parameter space is to be 
explored.  Therefore it is important to check its accuracy.  
Figure~\ref{exact_Limberfig} presents the relative error between the Limber 
calculation of kSZ power using~(\ref{Limberri}) and the exact calculation 
using~(\ref{Clexact}),
\beq
\fr{\Del C_\ell}{C_\ell^{\rm exact}}
   \equiv \fr{C_\ell^{\rm Limber} - C_\ell^{\rm exact}}{C_\ell^{\rm exact}},
\eeq
for the fiducial LTB model (which is described below).  The Limber 
approximation can be seen to be 
very accurate, to better than a tenth of a percent, over a wide range of 
angular scales.  This was expected, 
as the integrand $F(r)D(r,k)$ in~(\ref{Clexact}) varies slowly over the 
width of a void.\footnote{A more rapidly varying LTB profile might result in a 
less accurate Limber approximation.}  Henceforth, all calculations will 
be performed using the Limber approximation, (\ref{Limberri}), or 
equivalently~(\ref{Limberr}).

\section{Behaviour of kSZ power in void models}
\label{kSZbehavsec}

   With the formalism for calculating the kSZ power in place, we can 
now procede to calculate spectra for specific LTB models, and to 
investigate where in $k$ and $z$ space the kSZ power is sourced.  
First, the LTB background must be specified.  This entails the specification 
of a radial profile $K(r)$, together with the {\em local} Hubble rate at 
the centre today, $H_0$.  Our {\em fiducial profile} for $K(r)$ is 
defined in the \hyperref[LTBsec]{Appendix}.  Our fiducial {\em model} 
is further specified by the parameters $\omml = 0.2$, $z_L = 0.5$, and 
$h_0 = 0.71$, where $\omml \equiv \Om^{\rm loc}_{\rm m}(z = 0)$ is the local 
density parameter at the centre today, and $z_L$ is a measure of the width 
of the profile in redshift.  As we will see in section~\ref{constraintsec}, 
the fiducial model is a very good fit to the supernova data.

   Then the kSZ power can be calculated using 
the Limber expressions~(\ref{Limberri}) or~(\ref{Limberr}).  Importantly, 
in evaluating the function $F(r)$ using expressions~(\ref{Fdef}) 
to~(\ref{dzdr}), all background quantities must be evaluated consistently 
using the specified LTB background.  The dipole $\beta(r(z))$ is also 
calculated consistently in the LTB model by propagating past-directed 
null rays radially inwards and outwards from the point $(t(z),r(z))$ 
on the past light cone of the central observer to an early time, 
and comparing the resulting redshifts as described in~\cite{mzs10}.  
The matter power spectrum in the LTB model is assumed to match that 
of the standard \lfrw\ model, and is calculated as described in 
section~\ref{matterPSsec}.

\begin{figure}\begin{center}
%[ht]
%\includegraphics[width=0.7\columnwidth]{kSZ_Lambda_power.eps}
\includegraphics[width=\columnwidth]{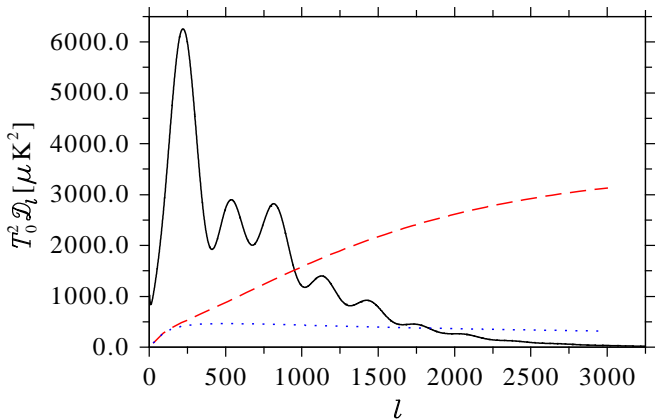}
\caption{The kSZ power calculated in the Limber approximation using the 
Halofit nonlinear matter power (dashed, red curve) and linear power 
(dotted, blue curve), compared with the temperature anisotropy power in 
the standard \lfrw\ model (solid, black curve).  Results are 
presented for the fiducial LTB model.
\label{kSZ_Lambda_powerfig}}
\end{center}\end{figure}

   A comparison of the kSZ power spectrum calculated for the fiducial LTB 
model with the temperature anisotropy power in the standard 
\lfrw\ model is presented in figure~\ref{kSZ_Lambda_powerfig}.  
The quantities plotted are the conventionally scaled power spectra,
\beq
{\cal D}_\ell \equiv \fr{\ell(\ell + 1)C_\ell}{2\pi}.
\eeq
Figure~\ref{kSZ_Lambda_powerfig} shows the kSZ power calculated using 
both the nonlinear (Halofit) matter power as well as the corresponding 
linear spectrum.  Nonlinearities in perturbation evolution enhance the matter 
power on small scales.  For the fiducial LTB model and the nonlinear matter 
power, the kSZ power is found to dominate over the standard temperature 
anisotropies for $\ell \gtrsim 1000$.  Clearly the presence of such 
a large kSZ component would result in a total anisotropy spectrum drastically 
different from that actually observed.  Although the kSZ power is reduced 
considerably in the linear matter power case, it still exceeds the 
standard temperature anisotropies by a large factor for the largest 
multipoles.  While the fiducial model would therefore appear to be 
immediately ruled out, we will see in section~\ref{varyparamsec} that 
the kSZ power depends sensitively on several parameters.

\begin{figure}\begin{center}
%[ht]
%\includegraphics[width=0.7\columnwidth]{dCldz.eps}
\includegraphics[width=\columnwidth]{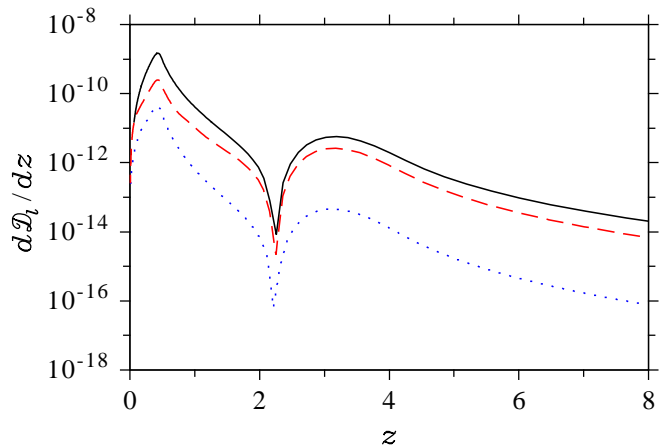}
\caption{The derivative of the kSZ power with respect to redshift, 
calculated in the Limber approximation using the Halofit nonlinear power 
for three multipoles: $\ell = 30$ (dotted, blue curve), $\ell = 300$ 
(dashed, red curve), and $\ell = 3000$ (solid, black curve).  Results are 
presented for the fiducial LTB model.
\label{dCldz_fig}}
\end{center}\end{figure}

   Figure~\ref{kSZ_Lambda_powerfig} suggests that the nonlinear matter 
power on small scales is the main source of the apparently very strong 
constraint that the kSZ can provide on void models.  We can examine this 
more explicitly by plotting derivatives of the kSZ power.  
Figure~\ref{dCldz_fig} shows the derivative with respect to redshift,
\beq
\fr{\rmd{\cal D}_\ell}{\rmd z} = \fr{8\pi\ell(\ell + 1)}{(2\ell + 1)^3}
          \fr{\rmd r}{\rmd z}rF^2(r)\Pd\ld(\fr{2\ell + 1}{2r},z(r)\rd).
\eeq
In words, it shows how the contributions to the total power at 
various distances along the line of sight are distributed.  The figure 
shows that, essentially independently of $\ell$, most of the kSZ power 
originates at redshifts $z \simeq 0.5$, \ie\ near the edge of the void 
for the fiducial model which extends to $z_L = 0.5$.  The redshift 
distribution of the kSZ power is largely determined by the dipole 
function, $\beta(z)$.  Indeed, the dipole peaks near $z = z_L$ and 
exhibits a zero at $z \simeq 2$ which is visible in figure~\ref{dCldz_fig}.

\begin{figure}\begin{center}
%[ht]
%\includegraphics[width=0.7\columnwidth]{dCldk.eps}
\includegraphics[width=\columnwidth]{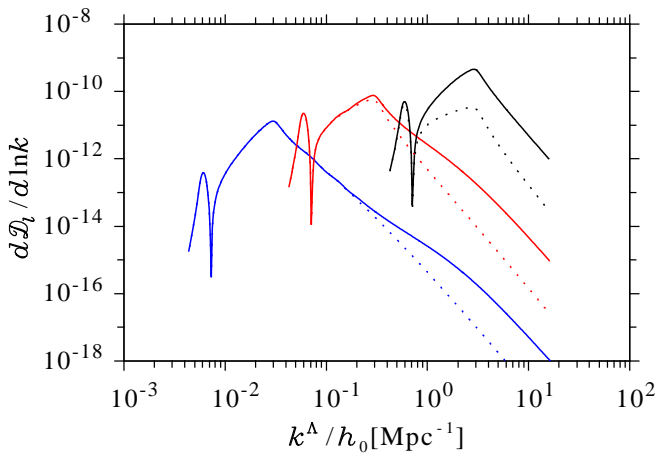}
\caption{The derivative of the kSZ power with respect to wave number, 
calculated in the Limber approximation using the Halofit nonlinear power 
(top, solid curve of each pair) and linear power (lower, dotted curves).  
Results are shown for three multipoles: $\ell = 30$ (leftmost, blue pair 
of curves), $\ell = 300$ (central, red curves), and $\ell = 3000$ 
(rightmost, black curves).  The distributions are plotted with respect to 
the scale $k^\Lam/h_0$, calculated as described in section~\ref{matterPSsec}, 
and measured as a proper wave number today.  Results are presented for the 
fiducial LTB model.
\label{dCldk_fig}}
\end{center}\end{figure}

  Figure~\ref{dCldk_fig} shows the derivative of kSZ power with respect to 
logarithmic $k$ interval,
\beq
\fr{\rmd{\cal D}_\ell}{\rmd\ln k} = \fr{2\pi\ell(\ell + 1)}{2\ell + 1}
          F^2\ld(\fr{2\ell + 1}{2k}\rd)\fr{\Pd(k,z)}{k^2},
\label{dDdk}
\eeq
for both the nonlinear and linear matter power spectrum.  That is, it 
shows how the contributions to the kSZ power from different length 
scales are distributed.  Importantly, the distributions are plotted 
with respect to the 
scale $k^\Lam$, calculated as described in section~\ref{matterPSsec} 
(and measured as a proper wave number today).  This is the scale that we 
would conclude a feature lies at if we mistakenly assumed that the \lfrw\ 
model was correct, when actually the specified LTB model was correct.

   As expected from the relation~(\ref{Deltatheta}), figure~\ref{dCldk_fig} 
shows that the wave numbers sourcing the kSZ power scale in inverse 
proportion to the angular scale examined on the sky.  In addition, it 
is clear that the length scales sourcing the most powerful part of the 
kSZ spectrum, near $\ell \simeq 3000$, are strongly nonlinear.  
Importantly, the figure also shows that the length scales we would infer 
to make the largest contribution to the high-$\ell$ kSZ effect are very 
small: for $\ell = 3000$, the corresponding length scales are roughly 
$1/3$ Mpc.  We will discuss the significance of these small scales in 
section~\ref{baryonsec}.

\section{Constraints on void models}
\label{constraintsec}

   The results of the previous section show that LTB models can 
potentially exhibit very high kSZ power.  In order to determine whether 
{\em viable} void models predict too much kSZ power, we must select 
models which fit other observations.  For our analysis we used the 
{\em minimal} amount of data required to 
fix the void profile, and hence obtained the most independent kSZ  
constraints possible.  To do this we used redshift-magnitude  
measurements of Type Ia SNe from the Union2 compilation~\cite 
{Amanullah:2010vv}, consisting of 557 SNe in the range $z = 0.015$--$1.4$.  
In our fitting we adopted the standard procedure of  
marginalizing over the unknown absolute magnitude, which is  
equivalent to marginalizing over $H_0$. A further input is therefore  
a {\em local} estimate of $H_0$ (\ie\ for $z \lesssim 0.1$), which is 
essentially free of any assumptions regarding the cosmological model.  
For this we used a prior of $h_0 = 0.72 \pm 0.08$ from the 
Hubble Space Telescope (HST) Key Project~\cite{Freedman:2000cf}.  Since our 
choice of prior has a larger error bar than the more recent measurements 
in~\cite{riessetal09,riessetal11}, it is more conservative.  Note that 
the SN data are not model independent; instead, they assume a standard 
\lfrw\ background (\eg, see the discussion in~\cite{ns11}).  This will 
result in further uncertainty in the kSZ power, considering the strong 
sensitivity of the power to the void width which we will explicate in 
section~\ref{varyparamsec}.

   In our previous analyses, we found that much lower values of the local 
Hubble rate ($h_0 \simeq 0.45$) were required for void models to fit the 
CMB primary anisotropies, assuming a near-power-law primordial 
spectrum~\cite{zms08,mzs10}.  Although such low values of $H_0$ are in 
severe conflict with the local measurements in~\cite{Freedman:2000cf,
riessetal09,riessetal11}, this tension might in principle 
be circumvented by substantial fine tuning of the primordial spectrum, 
allowing for higher values of $H_0$.  For this reason we chose a prior on 
$H_0$ which is consistent with local observations, and we did not include the 
CMB primaries in our constraints.  Another reason we excluded CMB primaries 
is that, as figure~\ref{kSZ_Lambda_powerfig} shows, a large linear kSZ 
component will substantially modify the observed anisotropies over a wide 
range of scales.

   We used \textsc{CosmoMC}~\cite{cosmomc} to generate 
Markov-Chain-Monte-Carlo 
chains to estimate confidence limits.  The LTB profile we used was the 
fiducial profile defined in the \hyperref[LTBsec]{Appendix}.  Three 
parameters were varied: the void width $L$ and depth $K_0$, along with 
$H_0$.  For each sample of the chain we computed several derived 
parameters: the redshift to $r = L$, $z_L$, the central matter density 
parameter, $\omml$, and the kSZ power, 
$\mathcal{D}_{\ell}$, at $\ell = 2500$, 3000, 3500, 4000, and 4500.  As in 
our previous analyses we applied a conservative prior of 
$\omml > 0.1$~\cite{carlbergetal97,fhp98}.  Deeper voids, 
with $\omml < 0.1$, must be correspondingly wider to fit 
the SN data and hence have the largest kSZ power (as we will see in 
figure~\ref{kSZ_power_Omz_fig}).  Hence the deepest voids are most 
strongly ruled out 
by the kSZ constraint.  We also applied a prior of $T_0^2 \mathcal{D}_
{\ell} < 10\:000\, \mu{\rm K}^2$, since such models would already be 
excluded at high significance by small-scale CMB experiments such as 
ACBAR~\cite{Reichardt:2008ay}.  Since we did not use primary CMB data, 
which fixes the baryon fraction $f_{\rm b}$ (at least at the radius of 
last scattering), we needed to 
choose a (local) value; we set the fiducial value of $f_{\rm b} = 0.168$ 
for the kSZ calculation.  Similarly, we set the fiducial value of 
$\sig_8 = 0.81$ for the matter power amplitude today.  The validity of 
these assumptions is discussed in the next section.

\begin{figure}\begin{center}
%[ht]
%\includegraphics[width=0.7\columnwidth]{kSZ_hist.eps}
\includegraphics[width=\columnwidth]{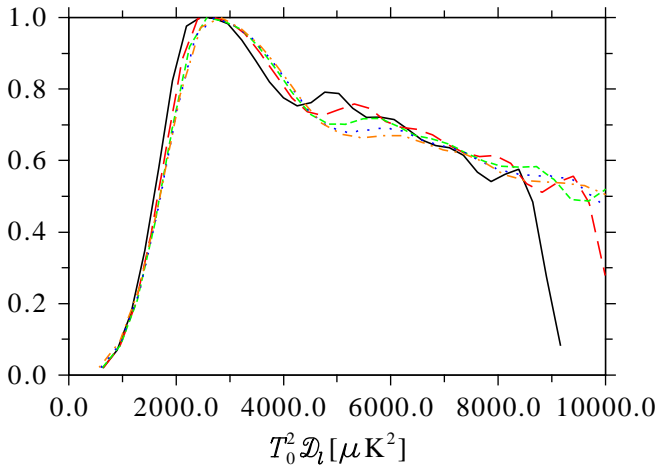}
\caption{Histograms of the kSZ power, calculated in the Limber 
approximation using the Halofit nonlinear power for five multipoles: 
$\ell = 2500$ (solid, black curve), $\ell = 3000$ (long dashed, red curve), 
$\ell = 3500$ (short dashed, green curve), $\ell = 4000$ (dotted, blue 
curve), and $\ell = 4500$ (dot-dashed, orange curve).  Results are 
presented for the fiducial LTB profile, fitted to the Union2 SN data, and 
using the HST prior of $h_0 = 0.72 \pm 0.08$.  Each curve is normalized to 
unity at its maximum.
\label{kSZ_hist_fig}}
\end{center}\end{figure}

   The best-fit model has a total $\chi^2 = 539$, and therefore a 
goodness-of-fit comparable to \lfrw\ and the void models in our 
previous work.  In figure~\ref{kSZ_hist_fig} we show a histogram of the 
kSZ power from the chains at 
various $\ell$ values. Each distribution is similar  over the $\ell$ 
range we consider, and hence lower limits on the kSZ power are 
largely independent of $\ell$.  The maximum-likelihood value is 
approximately $2500 \, \mu{\rm K}^2$, with an extended tail due to 
deeper, wider void profiles.  Asymmetric distributions make 
estimating confidence limits (from counting samples in the 
histogram) dependent on the tail, but we have checked the robustness 
of our constraints by also excluding samples with $T_0^2 \mathcal{D}_
{\ell} > 5000\, \mu{\rm K}^2$.  The $2\sigma$ lower limit on the kSZ 
power at $\ell = 3000$ is reduced from $1630\,\mu{\rm K}^2$ to 
$1440\,\mu{\rm K}^2$ when applying this constraint.  We take the lower (and 
more conservative) limit as our baseline value to compare with 
recent CMB observations on very small scales from the South Pole Telescope 
(SPT)~\cite{Hall:2009rv,shirokoffetal10} and the Atacama Cosmology Telescope 
(ACT)~\cite{Das:2010ga}.  In figure~\ref{kSZ_power_Omz_fig} we show the 
1, 2, and $3\sig$ confidence levels in the $z_L$-$\omml$ plane.  Note 
that our fiducial model, with $z_L = 0.5$ and $\omml = 0.2$, is very 
close to the best fit model.

   Single frequency bandpowers from the 150 GHz and and 148 GHz channels 
of SPT and ACT limit the {\em total} power (including primary CMB, 
point sources, thermal, and kinetic SZ) at $\ell=3000$ to $\lesssim 
50 \, \mu{\rm K}^2$.  While both SPT and ACT also present considerably 
tighter upper limits on the kinetic 
SZ component itself (and these limits were used in ZS10), such limits are 
model dependent.  This is because the contribution of primary CMB in 
particular may be quite different in LTB models than in \lfrw, recalling 
that the CMB and $H_0$ cannot be simultaneously fit without substantially 
modifying the primordial spectrum.  Therefore we consider the SPT and 
ACT total power to be a conservative upper limit to the kSZ power from 
voids.\footnote{The total powers will almost certainly be overestimates 
of the kSZ component, since they include a frequency-dependent component 
presumably due to point sources.}  Even taking this conservative approach, 
the predicted kSZ signal is at least a factor of roughly $30$ higher than 
the observational upper limits, and hence rules out void models which fit 
the SN data at high significance when the fiducial parameters are assumed.

\section{Robustness of kSZ constraints}
\label{robustsec}
\subsection{Parameter dependencies}
\label{varyparamsec}

   Although the results in section~\ref{constraintsec} suggest that the 
kSZ anisotropy power is large enough to immediately rule out void models 
for acceleration, it is important to examine the robustness of the kSZ 
power as a test of homogeneity.  In particular, how sensitive are the 
predictions to various model parameters?  This was already addressed 
to some extent implicitly in section~\ref{constraintsec}, where the LTB 
profile (width and depth), as well as local Hubble rate, were varied.  In 
this section we will illustrate these dependencies explicitly, and examine 
the effects of the remaining parameters.

\begin{figure}\begin{center}
%[ht]
%\includegraphics[width=0.7\columnwidth]{Omm_zL_kSZ_likely.eps}
\includegraphics[width=\columnwidth]{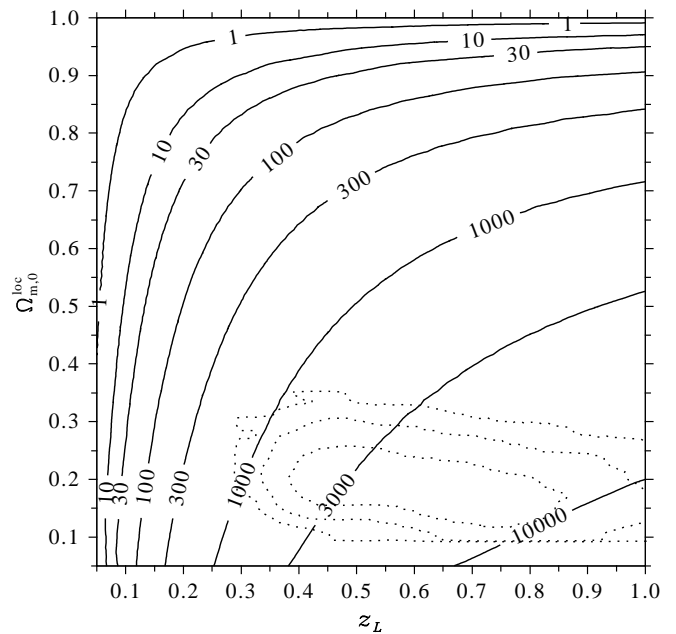}
\caption{Solid contours are the kSZ power, $T_0^2{\cal D}_{3000}$ 
(in $\mu{\rm K}^2$), calculated in the Limber approximation using the Halofit 
matter power, as a function of central matter density parameter, $\omml$, 
and a measure of the width of the void, $z_L$.  Results are presented for 
the fiducial LTB profile with $h_0 = 0.71$.  Dotted contours are the 1, 
2, and $3\sig$ confidence levels from the SN data, which used the HST prior 
of $h_0 = 0.72 \pm 0.08$.
\label{kSZ_power_Omz_fig}}
\end{center}\end{figure}

   Figure~\ref{kSZ_power_Omz_fig} shows the kSZ power for $\ell = 3000$ 
as a function of a measure of the width of the void, $z_L$, and the 
depth of the void as measured by the central density parameter, $\omml$.  
The LTB profile used was the fiducial profile, with local Hubble rate 
$h_0 = 0.71$.  Superimposed on figure~\ref{kSZ_power_Omz_fig} are the 1, 
2, and $3\sig$ confidence levels from the Union2 SN data calculated in 
section~\ref{constraintsec}.  The kSZ power is seen to be a very 
sensitive function of the void width, varying by roughly a decade over 
the region allowed by the SNe.  Note that, for the fiducial profile, the 
multi-valued region of parameter space (for which $H_\parallel(z) \le 0$) 
occurs only for the deepest voids, with $\omml < 0.05$.  Therefore, as 
noted in section~\ref{matterPSsec}, the kSZ calculation is not valid in 
this regime and we do not plot it in figure~\ref{kSZ_power_Omz_fig}.  
(The excluded shell-crossing region occurs for even deeper voids than the 
multi-valued models.)

   We can understand the general forms of the parameter dependencies 
illustrated in figure~\ref{kSZ_power_Omz_fig} quite easily.  For 
fixed void width $z_L$ (and fixed $H_0$), the kSZ power increases as 
$\omml$ decreases from unity.  This is mainly due to 
the dipole factor $\beta^2$ in the expression for the kSZ, 
(\ref{Limberri}).  Inside the void, we can define an effective curvature 
parameter by $\Om^{\rm eff}_K \equiv 1 - \Om^{\rm loc}_{\rm m}$.  Then, 
for a growing mode background solution, we expect $\Om^{\rm eff}_{K,0} 
\propto \del H/H$, where $\del H$ is some measure of the perturbation 
in expansion rate inside the void relative to the outside~\cite{z08}.  
Finally, we expect the dipole at fixed redshift to be proportional to 
the expansion perturbation,
\beq
\beta \simeq D\del H,
\label{betadist}
\eeq
where $D$ is some proper distance measure~\cite{aa06}.  Combining these 
relations we find 
\beq
C_\ell \propto \beta^2(z_L) \propto \ld(1 - \omml\rd)^2.
\eeq
Although there are considerable ambiguities in defining most of the 
quantities used to derive this relationship, it is satisfied reasonably 
well by the numerical results of figure~\ref{kSZ_power_Omz_fig}.

   Next, for fixed $\omml$ (and fixed $H_0$), 
figure~\ref{kSZ_power_Omz_fig} shows that the kSZ power grows rapidly with 
increasing void width $z_L$.  Using the relation~(\ref{betadist}), we 
have $\beta^2(r) \propto r^2$, since coordinate $r$ will be approximately 
proportional to a proper distance measure.  Then, ignoring the scale 
dependence of the matter power spectrum, the kSZ power 
expression~(\ref{Limberri}) implies
\beq
C_\ell \propto \int_0^L r^3\rmd r \propto L^4 \propto z_L^4,
\eeq
for sufficiently small $z_L$.  Again, this relation is indeed roughly 
satisfied by the numerical results.

\begin{figure}\begin{center}
%[ht]
%\includegraphics[width=0.7\columnwidth]{H0_zL_kSZ.eps}
\includegraphics[width=\columnwidth]{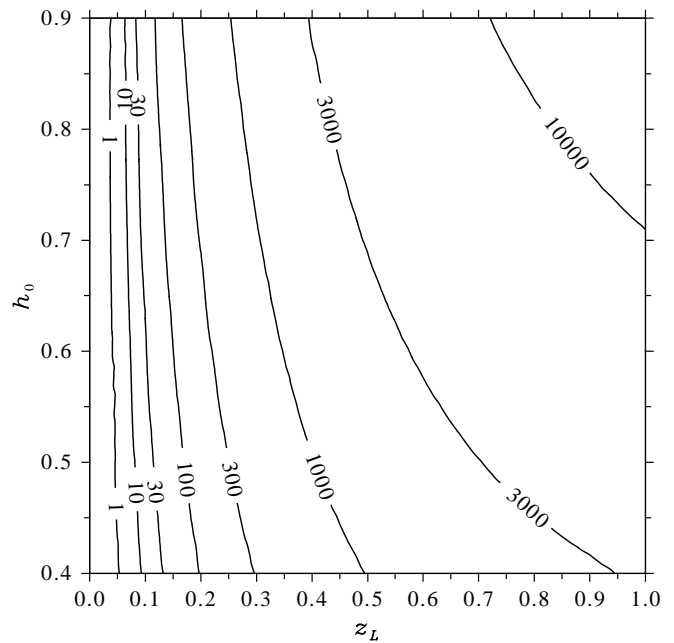}
\caption{The kSZ power, $T_0^2{\cal D}_{3000}$ (in $\mu{\rm K}^2$), 
calculated in the Limber approximation using the Halofit matter power, 
as a function of local Hubble rate, $H_0$, and a measure of the width 
of the void, $z_L$.  Results are presented for the fiducial LTB profile 
with $\omml = 0.2$.
\label{power_H_fig}}
\end{center}\end{figure}

   Figure~\ref{power_H_fig} illustrates the effect of varying the local 
Hubble rate, $H_0$, while keeping the LTB depth (as measured by $\omml$) 
fixed.  Larger $H_0$ leads to larger kSZ power, and this dependence is 
due to a combination of background factors in the relations~(\ref{Fdef}) 
to~(\ref{dzdr}), together with the scale dependence of the matter power.  
The variation in power over our chosen prior range of $h_0 = 0.72 \pm 
0.08$~\cite{Freedman:2000cf} is not large, and of course the variation is 
considerably smaller over the range of the newer measurement, 
$h_0 = 0.738 \pm 0.024$~\cite{riessetal11}.

   The remaining parameters that affect the kSZ anisotropy power are 
the linear matter power amplitude today, $\sig_8$, conventionally expressed 
on a scale of $8/h_0$ Mpc, and the baryon fraction $f_{\rm b}$.  The 
fluctuation amplitude could also be expressed in terms of the primordial 
perturbation amplitude, although in our approach of specifying the 
matter power at late times rather than evolving perturbation ICs, 
the parameter $\sig_8$ is more natural.  If the fluctuations were 
{\em linear,} the kSZ power would simply scale as $\sig_8^2$.  However, 
in the nonlinear regime relevant to the kSZ effect for high $\ell$, 
we expect~\cite{dhs04}\footnote{For the ordinary (nonlinear) kSZ effect 
in standard \lfrw\ models, the kSZ power scales like an even larger power 
of $\sig_8$, since the same amplitude also determines the velocity 
perturbation spectrum~\cite{dhs04}.}
\beq
C_\ell \propto \Pd \propto \sig_8^{2-3}.
\eeq
This means that, \eg, an uncertainty of 30\% in $\sig_8$ would lead to an 
uncertainty in the kSZ power of roughly a factor of two.

   Of course, observations of the CMB are usually considered 
to constrain $\sig_8$ very tightly, to a few percent~\cite{wmap7params}.  
But this is based on the assumption of homogeneity: void models, on the 
other hand, contain an extremely large inhomogeneity of mysterious 
origin, and hence there is no necessary link between the perturbation 
amplitude locally and that at the last scattering radius.  In addition, as 
mentioned in the \hyperref[introsec]{Introduction}, if we are to satisfy 
local measurements of $H_0$, the primordial spectrum must be modified 
substantially~\cite{zms08,mzs10}, further clouding the connection between 
the local and CMB amplitudes.  {\em Local} measurements of $\sig_8$ 
(or, more precisely, measurements at $z \simeq 0.5$) 
are what we need to test void models via the kSZ effect; unfortunately, 
they are considerably more uncertain than those 
based on the CMB.  For example, measurements based on weak lensing 
(\eg,~\cite{fuetal08}) constrain only a combination of $\sig_8$ and 
$\Om_{\rm m}$.  Extracting constraints on $\sig_8$ from these results 
would likely be strongly model dependent.  Recalling that our calculations 
have used the standard CMB-based value, $\sig_8 = 0.81$~\cite{wmap7params}, 
we can expect the uncertain local matter normalization to add significant 
ambiguity to our calculations.

  The final parameter determining the kSZ power is the baryon fraction 
$f_{\rm b}$.  This affects the kSZ via the background optical depth, 
which is related to the background electron density (assumed equal to 
the baryon density) through~(\ref{dtaudz}).  A {\em constant} baryon 
fraction affects the kSZ power through the simple scaling
\beq
C_\ell \propto f_{\rm b}^2.
\eeq
As was the case with the matter power amplitude, constraints on $f_{\rm b}$ 
from the CMB will not generally be applicable in an inhomogeneous 
Universe, and, again, local measurements are not very 
constraining.  In addition, in void models, the presence of 
cosmological-scale adiabatic (curvature) inhomogeneity of mysterious 
origin suggests that we should also consider the possibility of isocurvature 
inhomogeneity between baryons and dark matter, in the form of an 
$r$-dependent $f_{\rm b}$~\cite{cr10}.  
Thus, recalling that in our calculations we have used the CMB-based value 
$f_{\rm b} = 0.168$~\cite{wmap7params}, our poor knowledge of the local 
baryon fraction in the context of LTB cosmologies will result in considerable 
further uncertainty in the kSZ power calculations presented here.

\subsection{Small-scale baryonic astrophysics}
\label{baryonsec}

   Recall from figure~\ref{dCldk_fig} that, for the large $\ell$'s which 
provide the strongest constraints on void models, the dominant length 
scales sourcing the kSZ in the fiducial model are very small, namely 
less than a Mpc.  Matter perturbations on such small scales are expected 
to be strongly nonlinear, as we have already seen in section~\ref{kSZbehavsec} 
when comparing the kSZ power calculated for linear and nonlinear 
matter power spectra.  However, when we began the derivation of the kSZ 
effect, we assumed that the free electron fluctuations match those of 
the total matter, so that $\del\rho_e/\rho_e = 
\del\rho_{\rm m}/\rho_{\rm m}$.  In the nonlinear regime, the baryonic 
(and hence electron) fluctuations are expected to be suppressed with 
respect to the total (or dark) matter power, due to interactions that 
become important on small scales (\eg, see~\cite{jzlgs06}).  Therefore, 
we expect the kSZ power sourced at these scales to be somewhat smaller 
than the calculations thus far (which used the standard \lfrw\ {\em total} 
matter power spectra) have indicated.  To help examine the importance of 
this effect, it will be useful to consider the dependence of the sourcing 
length scales on the various model parameters.

\begin{figure}\begin{center}
%[ht]
%\includegraphics[width=0.7\columnwidth]{Omm_zL_kmax_likely.eps}
\includegraphics[width=\columnwidth]{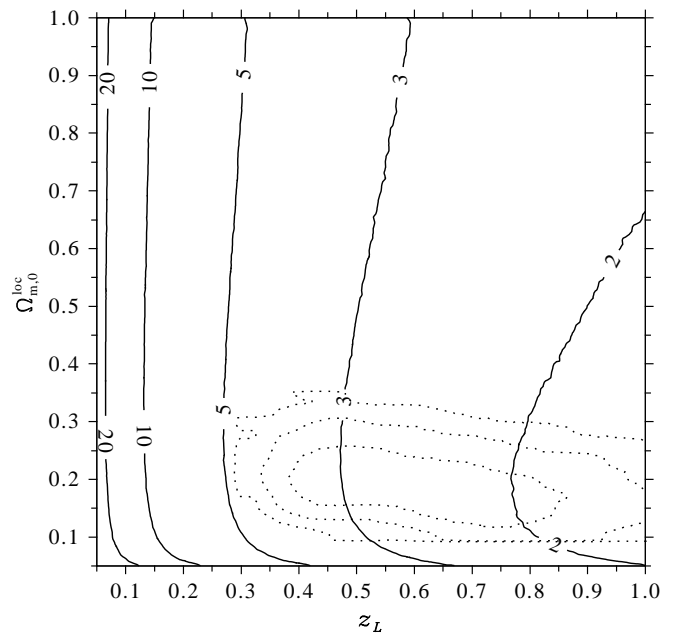}
\caption{Solid contours are the peak wave number contributing to the kSZ 
power at $\ell = 3000$, 
calculated in the Limber approximation using the Halofit nonlinear power, 
as a function of central matter density parameter, $\omml$, 
and a measure of the width of the void, $z_L$.  The contour values are the 
peak scale $k^\Lam/h_0$ (in Mpc$^{-1}$), calculated as described in 
section~\ref{matterPSsec}, and measured as a proper wave number today.  
Results are presented for the fiducial LTB profile with $h_0 = 0.71$.  
Dotted contours are the 1, 2, and $3\sig$ confidence levels from the SN 
data, which used the HST prior of $h_0 = 0.72 \pm 0.08$.
\label{kmax_Omz_fig}}
\end{center}\end{figure}

   In figure~\ref{kmax_Omz_fig} we illustrate the dependence on the 
width and depth of the void of the wave number providing the peak 
contribution to the $\ell = 3000$ kSZ power, again for the fiducial 
profile and $h_0 = 0.71$.  (The plotted peak wave number maximizes 
$\rmd{\cal D}_\ell/\rmd\ln k$, given in (\ref{dDdk}).)  As we did for 
figure~\ref{dCldk_fig}, we 
here plot the $k$ scale that we would {\em conclude} sources the kSZ, 
if we mistakenly assumed that the correct model was standard \lfrw, 
when actually the (fiducial) void model was correct.  This effective 
\lfrw\ scale is calculated according to the prescription of 
section~\ref{matterPSsec}.  Figure~\ref{kmax_Omz_fig} shows that the peak 
$k$ scale generally increases as the void width decreases.  This can be 
understood simply from the geometrical relation~(\ref{Deltatheta}): 
keeping $\ell$ (or the angular scale) fixed, the relevant $k$ scale 
must be inversely proportional to $z$, since $d_{\rm A}(z) \propto z$, 
for sufficiently small redshift.  This simple relationship is modified 
for large $z$ or very deep voids, since relativistic effects then become 
important.

   For the entire region of parameter space illustrated in 
figure~\ref{kmax_Omz_fig}, the peak scale is smaller than 1 Mpc$^{-1}$, 
and for the narrowest voids consistent with the SN constraints from 
section~\ref{constraintsec}, namely $z_L \simeq 0.35$ at $2\sig$, the 
peak scale is $k^\Lam/h_0 \simeq 4\,{\rm Mpc}^{-1}$.  
On such small scales, hydrodynamical simulations indicate considerable 
suppression of baryon power with respect to dark matter power, although 
the various studies differ quantitatively in their predictions 
(\eg, see~\cite{dst05,jzlgs06}).  Observations, on the other hand, most 
directly probe either the galaxy or the total matter power, rather 
than the baryon power.  Therefore, it is clear that the poorly 
understood details of small-scale baryonic astrophysics add further 
considerable uncertainty to our estimates of kSZ power, although in 
this case we can say that our calculations will {\em over}estimate 
the kSZ for large $\ell$'s.  It is also clear that the models most 
likely to avoid producing too much kSZ power will be the narrowest 
profiles consistent with the SN data: narrower profiles both produce 
less kSZ power when baryon suppression is ignored, and also result in 
greater baryon suppression, since the relevant length scales are the smallest.

\section{Importance of the consistent LTB framework}
\label{imptltbsec}

   At first glance, the kSZ power values in the $z_L$-$\omml$ plane 
presented in figure~\ref{kSZ_power_Omz_fig} appear remarkably 
similar to those of ZS10, considering the fully relativistic treatment 
of the background and the detailed treatment of the $k$ scales performed 
here.  However, note that for generic void profiles there is 
considerable arbitrariness in attempting to define an ``edge'', with 
corresponding redshift $z_L$.  Therefore, we only expect plots of this 
kind for different profiles to agree up to a scaling of 
the $z_L$ axis by a factor of roughly unity.  Considering the strong 
sensitivity of kSZ power to $z_L$ exhibited in figure~\ref{kSZ_power_Omz_fig}, 
this arbitrariness means that we can only say that our kSZ values are 
in agreement with those of ZS10 to within roughly an order of magnitude.  
This point is particularly relevant considering how different our smooth, 
polynomial fiducial LTB profile is from the ``step function'' Hubble bubble 
profile used in ZS10.

   While we cannot state how precisely our kSZ power values agree with 
those of ZS10, we {\em can} directly compare our kSZ values for models which 
satisfy the SN constraints.  Superimposed on figure~\ref{kSZ_power_Omz_fig} 
are the 1, 2, and $3\sig$ confidence levels from the Union2 SN data calculated 
in section~\ref{constraintsec}.  Comparing these contours with those in 
ZS10, we can see that the kSZ power values for models which satisfy the SN 
constraints at $2\sig$ are roughly {\em one tenth} those found in ZS10.  
A related observation is that ZS10 quote a minimum $\chi^2$ of $605.4$ 
for their void models with respect to the 557 Union2 SNe, which represents 
a much worse fit than our best-fit value of $\chi^2 = 539$.  We believe that 
the poorness of fit of the SZ10 profiles to the SNe accounts for most of 
the roughly tenfold difference in our results.  This illustrates the 
importance of performing the kSZ calculations 
within a consistent, LTB framework, with smooth radial profiles which 
provide a good fit to the SNe.  To obtain meaningful estimates of the kSZ 
power in void models, we cannot rely on void profiles which provide a poor 
fit to the SNe.

   We can reinforce this point by calculating a quantitity related to, but 
distinct from, the kSZ power, namely the Compton $y$-distortion of the 
CMB frequency spectrum.  This distortion arises from the scattering of 
CMB photons from inside our past light cone into our line of sight, and 
has been used to provide constraints on void 
models~\cite{goodman95,cs08,mzs10}.  Like the kSZ effect, the $y$-distortion 
also depends on the dipole $\beta(z)$ along our past light cone, but is a 
background-level effect, so is independent of the matter power spectrum.  
In the single-scattering and linear approximations, and when the dipole 
anisotropy dominates, the $y$-distortion can be written as~\cite{mzs10}
\beq
y = \fr{7}{10}\int_0^{r_{\rm re}} \rmd r
              \fr{\rmd\tau}{\rmd z}\fr{\rmd z}{\rmd r}\beta(r)^2
  = \fr{7}{10}\int_0^{r_{\rm re}} \rmd r F(r)\beta(r),
\label{ydistn}
\eeq
where $r_{\rm re}$ is the radial coordinate of (the assumed abrupt) 
reionization.

\begin{figure}\begin{center}
%[ht]
%\includegraphics[width=0.7\columnwidth]{Omm_zL_ydistn_likely.eps}
\includegraphics[width=\columnwidth]{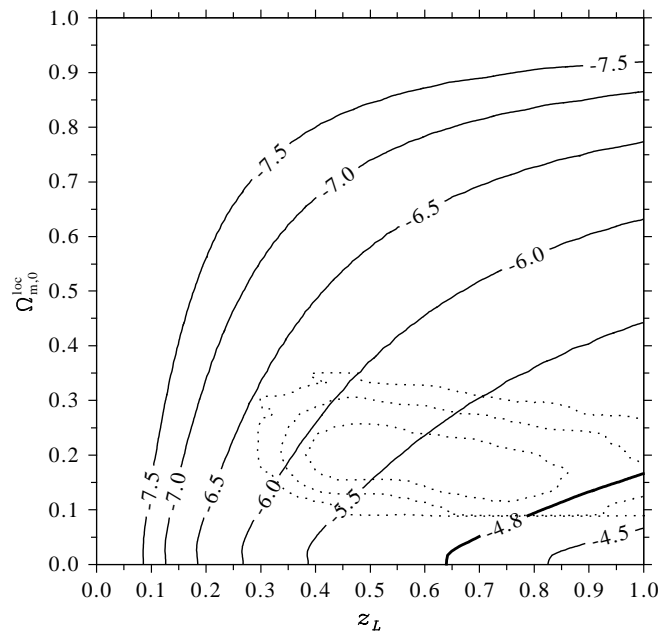}
\caption{Solid contours are the $y$-distortion, as a function of central 
matter density parameter, $\omml$, and a measure of the width of the void, 
$z_L$.  The contour values are $\log_{10}y$, calculated 
using~(\ref{ydistn}), and the heavy contour represents the COBE upper limit 
of $y < 1.5\times10^{-5}$~\cite{fixsenetal96}.  Results are presented for 
the fiducial LTB profile with $h_0 = 0.71$.  Dotted contours are the 1, 2, 
and $3\sig$ confidence levels from the SN data, which used the HST prior 
of $h_0 = 0.72 \pm 0.08$.
\label{ydistn_Omz_fig}}
\end{center}\end{figure}

   The similarity of expression~(\ref{ydistn}) to expression~(\ref{Limberr}) 
for the kSZ power in the Limber approximation means that we can readily 
calculate the $y$-distortion for our fiducial profile.  In 
figure~\ref{ydistn_Omz_fig} we plot the $y$-distortion in the $z_L$-$\omml$ 
plane for the fiducial LTB profile.  The heavy contour indicates the 
$2\sigma$ upper limit of $y < 1.5\times10^{-5}$ from the COBE 
satellite~\cite{fixsenetal96}.  Superimposed on the figure are our 1, 2, 
and $3\sig$ confidence levels from the Union2 SN data.  It is apparent 
that almost all of the models allowed by the SN data have $y$-distortion 
values below the COBE limit, and hence the $y$-distortion provides no 
significant constraint on void models.

   This result is in sharp contrast to~\cite{cs08}, who found that almost 
all models allowed by SN data were {\em above} the COBE limit, and hence 
ruled out.  Importantly, the void models used in~\cite{cs08} were the 
same Hubble bubble models as were studied in ZS10.  The simplicity of the 
$y$-distortion calculation means that it is free of the ambiguities of 
harmonic decomposition, $k$ scales, and small-scale baryonic astrophysics 
which plague the kSZ calculation.  Therefore, this result demonstrates 
again that the treatment of the background spacetime, and, in particular, 
obtaining good fits to the SN data, is crucial in obtaining meaningful 
constraints on void models for acceleration.

\section{Conclusions}
\label{conclsec}

   Our results indicate that void models which satisfy local constraints 
on the Hubble rate and fit the SN observations predict considerably 
higher kSZ power for $\ell \simeq 2000$--$3000$ than the {\em total} 
anisotropy power observed at those scales by SPT and ACT.  Thus 
all local void models would appear to be ruled out.  To evade these 
high-$\ell$ constraints, the kSZ power would need to be reduced by a 
factor of roughly 30 for the narrowest voids allowed by the SN data, 
with our conservative interpretation of the power measurements.  We 
have highlighted several caveats and ambiguities that render our 
calculations uncertain and could in principle lead to such a reduction.  
By way of summary, we list them here:
\renewcommand{\labelenumi}{(\arabic{enumi})}
\begin{enumerate}
\item We have needed a harmonic decomposition of the matter fluctuations, 
although it is unclear how this should be done at late times on an LTB 
background.
\item We have attempted to match the late time matter power in 
the LTB model to that in standard \lfrw\ via the physically correct 
prescription, but we have found that it is not possible to do this 
without ambiguity in the $k$ scales and hence in the matter power.
\item We have assumed that the linear kSZ contribution 
$\int\del\beta\rmd\tau$ vanishes in LTB models as it does in FLRW, 
although this would need to be checked explicitly with a proper LTB 
perturbation formalism.
\item The kSZ power is very sensitive to the parameters $\sig_8$ and 
$f_{\rm b}$, but their {\em local} values are very uncertain, both 
theoretically and observationally.
\item We have shown that the kSZ power for large $\ell$ is sourced mainly 
by very small scales, between 1 and $0.1$ Mpc.  The baryon power is 
expected to be suppressed significantly relative to total matter power on 
these scales, reducing the kSZ power by an uncertain amount.
\item The strong sensitivity of kSZ power to void width means that model 
dependence in the SN data, or simply changes to the void radial profile, 
may substantially affect the kSZ power.
\item It is likely that the dipole values along the light cone, and hence 
the kSZ power, could be reduced by choosing an appropriate isocurvature 
profile between radiation and matter at early times~\cite{cr10,yns10b}, 
although it is likely that this would entail fine tuning.
\item It is also possible that including a significant late-time LTB 
decaying mode could evade the kSZ constraints, although this would entail 
a drastic departure from the standard view of the early Universe.
\end{enumerate}
To determine the importance of the first three points listed above would 
require a proper treatment of perturbations on LTB backgrounds.  To reduce 
the uncertainty due to points (4) and (5) would require a direct 
measurement of the {\em baryon} power spectrum on $\sim$$1/3$ Mpc scales, 
as well as a measurement of the baryon fraction, both at redshifts 
$z \simeq 0.5$.  It is clear, though, that a void must be as narrow as 
possible if it is to minimize the kSZ power.

   Our result that void models with the fiducial parameters predict, at 
$2\sig$ confidence, at least a factor of roughly 30 larger kSZ power 
than that observed is considerably weaker than the corresponding result 
of ZS10, who found a discrepancy of more than three orders of magnitude 
for voids that fit the SNe at $2\sig$ confidence.  The difference 
between our results is presumably mainly due to our rigorous LTB 
treatment of the background together with our careful treatment of wave 
numbers.  Our more conservative interpretation of the SPT and ACT power 
measurements also contributes to the difference in our results.  This 
conclusion is reinforced by the $y$-distortion values for our models, which 
are strongly at odds with those of~\cite{cs08}, who also used a simple 
Hubble bubble model.  Again, this emphasizes the importance of using smooth 
radial profiles which fit the SN data well in order to obtain meaningful 
constraints on void models.

   As we stressed in the \hyperref[introsec]{Introduction}, if a void 
model is to generate the observed primary CMB anisotropies, it must have a 
{\em local} Hubble rate low enough ($h_0 \simeq 0.45$) to rule out all such 
models~\cite{zms08,mzs10,bnv10}.  This conclusion is independent of the 
poorly understood details of the IC's or evolution of perturbations on LTB 
backgrounds, and it is insensitive to the uncertainties of small-scale 
baryonic astrophysics as well as to the values of locally poorly 
determined parameters such as $\sig_8$ and $f_{\rm b}$.  For growing 
mode LTB profiles, the only chance to avoid this CMB constraint is 
possibly by substantially altering the primordial spectrum from scale 
invariance~\cite{mzs10}.  But it is not enough, of course, to simply 
introduce a significant non-scale-invariant feature: that feature must be 
just so contrived that together with the local void profile it {\em mimics} 
the near-scale-invariant spectrum expected in the standard \lfrw\ model!  
In~\cite{ns11}, \eg, a five-parameter modification of the primordial 
spectrum was required in order to match only the WMAP CMB data (no small-scale 
CMB data was used), with the somewhat improved local Hubble rate of 
$h_0 \simeq 0.60$.  The effect of an early radiation inhomogeneity~\cite{cr10} 
should be negligible at late times~\cite{mzs10}, and hence it appears 
unlikely that this could provide a loophole.

   The superb sensitivity of the linear kSZ approach goes a very substantial 
way towards closing the rather contrived loophole of modifying the 
primordial spectrum for void models.  However, the various ambiguities 
in its application to void models should be considerably less important 
in a related but distinct application: constraining departures from 
homogeneity within the standard \lfrw\ framework.  That is, {\em assuming} 
that the background is near-FLRW with the usual fraction of dark energy, 
and that the parameters $\sig_8$ and $f_{\rm b}$ are as indicated by the 
CMB, we can use the linear kSZ effect to constrain small departures from 
homogeneity, as done in ZS10.  Also, at greater redshifts, the length 
scales sourcing the kSZ at fixed $\ell$ will be larger, reducing the 
uncertainty in baryonic astrophysics.  Thus the linear kSZ effect should 
bring us closer to observationally confirming the long-held belief in 
cosmological homogeneity.

\begin{acknowledgments}
%\ack
   We thank Douglas Scott and Albert Stebbins for useful discussions.  This 
research was supported by the Canadian Space Agency.
\end{acknowledgments}

%\appendix
%\section*{Appendix.  The Lema\^itre-Tolman-Bondi solution}
%\setcounter{section}{1}
\appendix*
\section{The Lema\^itre-Tolman-Bondi solution}
\label{LTBsec}

   We model the Universe at background level as a spherically symmetric 
distribution of pressureless matter, with the observer at the centre.  
For this spacetime, Einstein's equations can be solved exactly, resulting 
in the LTB solution.  The metric can be written
\beq
\rmd s^2 = -\rmd t^2 + \fr{Y'^2}{1 - K}\rmd r^2 + Y^2\rmd \Omega^2,
\eeq
where a prime denotes the derivative with respect to comoving radial 
coordinate $r$, and $t$ is the proper time along the comoving worldlines.  
The function $K = K(r)$ is arbitrary, and the areal radius $Y = Y(t,r)$ 
is given parametrically by
%\bea
%Y &= \fr{M}{K}(1 - \cosh\eta), \quad
%t - t_{\rm B} = \fr{M}{(-K)^{3/2}}(\sinh\eta - \eta), \qquad&K < 0,\\
%Y &= \fr{M}{K}(1 - \cos \eta), \quad
%t - t_{\rm B} = \fr{M}{K^{3/2}}(\eta - \sin\eta ),          &0 < K < 1,\\
%Y &= \ld(\fr{9M}{2}\rd)^{1/3}\ld(t - t_{\rm B}\rd)^{2/3},   &K = 0.
%\eea
%\bea
%Y &=&  \ld\{\begin{array}{ll}
%&{\displaystyle\fr{M}{K}(1 - \cosh\eta)}\\[0.2cm]
%&{\displaystyle\fr{M}{K}(1 - \cos \eta)}\\
%    \end{array}\rd.,
%\quad
%t - t_{\rm B} = \ld\{\begin{array}{ll}
%{\displaystyle\fr{M}{(-K)^{3/2}}(\sinh\eta - \eta)}\qquad &K < 0,\\[0.2cm]
%{\displaystyle\fr{M}{K^{3/2}}(\eta - \sin\eta )}          &0 < K < 1,\\
%    \end{array}\rd.\\
%Y &=& \ld(\fr{9M}{2}\rd)^{1/3}\ld(t - t_{\rm B}\rd)^{2/3},    K = 0.
%\eea
\beq
Y = \ld\{\begin{array}{ll}
{\displaystyle\fr{M}{K}(1 - \cosh\eta)}            &K < 0,\\[0.3cm]
{\displaystyle\fr{M}{K}(1 - \cos \eta)}            &0 < K < 1,\\[0.3cm]
{\displaystyle\ld(\fr{9M}{2}\rd)^{1/3}\ld(t-t_{\rm B}\rd)^{2/3}}\quad&K = 0,\\
       \end{array}\rd.
\eeq
\beq
t - t_{\rm B} = \ld\{\begin{array}{ll}
{\displaystyle\fr{M}{(-K)^{3/2}}(\sinh\eta - \eta)}\quad &K < 0,\\[0.3cm]
{\displaystyle\fr{M}{K^{3/2}}(\eta - \sin\eta )}         &0 < K < 1.\\
                \end{array}\rd.\\
\eeq
This exact solution reveals the presence of a second free radial function, 
$t_{\rm B} = t_{\rm B}(r)$, which is known as the ``bang time'' function, 
since the cosmological singularity occurs at $t = t_{\rm B}(r)$.  Throughout 
this work we assume a homogeneous big bang and set $t_{\rm B} = 0$.  In this 
case the LTB solution contains no decaying mode~\cite{silk77,z08}, 
and hence we retain the standard inflationary picture of an essentially 
homogeneous early (but post-inflation) Universe.  There is also a third free 
radial function, $M(r)$, which we set to $M(r)= r^3$ without loss of 
generality as our gauge condition.

   The radial and transverse comoving expansion rates are given by 
$H_\parallel = {\dot Y}'/Y'$ and $H_\perp = {\dot Y}/Y$, respectively, 
where the overdot denotes the derivative with respect to $t$.  We define 
the local density parameter by $\Om^{\rm loc}_{\rm m} \equiv 24\pi 
G\rho_{\rm m}/\theta^2$, where $\rho_{\rm m}$ is the {\em total} matter 
density and $\theta$ the comoving volume expansion.  At the centre of symmetry 
in an arbitrary LTB spacetime, we have $H_\parallel = H_\perp \equiv H$.  
The angular diameter distance from the centre to redshift $z$ is simply 
$d_{\rm A} = Y(z)$.

   The LTB spacetime is completely determined once the single free 
radial profile $K(r)$ is specified.  Based on our previous experience with 
a wide range of LTB profiles~\cite{zms08,mzs10}, we chose for this study 
a simple profile which depends on only two parameters, a width and a depth.  
Explicitly, we chose
%\bea
%K(r) &= K_0\ld[\ld(\fr{r}{L}\rd)^5 - \fr{9}{5}\ld(\fr{r}{L}\rd)^4
%         + \ld(\fr{r}{L}\rd)^2\rd], \qquad&r \le L,\label{K1}\\
%K(r) &= \fr{K_0}{5}\fr{L}{r},             &r > L.\label{K2}
%\eea
\beq
K(r) = \ld\{\begin{array}{ll}
       {\displaystyle K_0\ld[\ld(\fr{r}{L}\rd)^5 - \fr{9}{5}\ld(\fr{r}{L}\rd)^4
         + \ld(\fr{r}{L}\rd)^2\rd]}    \quad&r \le L,\label{K1}\\[0.3cm]
       {\displaystyle \fr{K_0}{5}\fr{L}{r}} &r > L.\label{K2}
    \end{array}\rd.
\eeq
The profile width and depth are determined by the parameters $L$ and $K_0$, 
respectively, but it is normally more physically relevant to specify the 
parameter $\omml \equiv \Om^{\rm loc}_{\rm m}(z = 0)$ at the observation 
point to characterize 
the depth, and to specify the redshift $z_L \equiv z(r = L)$ to determine 
the width.  Finally, the observation point is determined by the corresponding 
proper time coordinate, $t_0$, or, more observationally relevantly, the 
Hubble rate today at the centre, $H_0 \equiv H(t_0)$.  We will refer to 
%the general profile, (\ref{K1}) and~(\ref{K2}), with unspecified $K_0$, $L$, 
the general profile, (\ref{K1}), with unspecified $K_0$, $L$, 
and $H_0$, as the {\em fiducial profile.}  The model with the specific values 
$\omml = 0.2$, $z_L = 0.5$, and $h_0 = 0.71$ (where 
$H_0 \equiv 100\, h_0 \, {\rm km} \, {\rm s^{-1}} \, {\rm Mpc^{-1}}$) will 
be referred to as the {\em fiducial model;} as shown in 
section~\ref{constraintsec}, this model is a very good fit to the supernova 
data.  In all of our calculations we chose the values $f_{\rm b} = 0.168$ and 
$\sig_8 = 0.81$ for the baryon fraction and matter amplitude today, as CMB 
observations suggest~\cite{wmap7params} in the context of FLRW models.  
In our numerical code, we check for the presence of shell crossing 
singularities along the past light cone, and exclude any models which 
exhibit them, since the LTB solution is not valid in such a case.  Further 
details and a discussion of our numerical implementation are provided 
in~\cite{mzs10}.

%\section*{References}
\bibliography{bib}

\begin{thebibliography}{61}
\expandafter\ifx\csname natexlab\endcsname\relax\def\natexlab#1{#1}\fi
\expandafter\ifx\csname bibnamefont\endcsname\relax
  \def\bibnamefont#1{#1}\fi
\expandafter\ifx\csname bibfnamefont\endcsname\relax
  \def\bibfnamefont#1{#1}\fi
\expandafter\ifx\csname citenamefont\endcsname\relax
  \def\citenamefont#1{#1}\fi
\expandafter\ifx\csname url\endcsname\relax
  \def\url#1{\texttt{#1}}\fi
\expandafter\ifx\csname urlprefix\endcsname\relax\def\urlprefix{URL }\fi
\providecommand{\bibinfo}[2]{#2}
\renewcommand{\eprint}[1]{arXiv:\href{http://arxiv.org/abs/#1}{#1}}
\providecommand{\doi}[2]{\href{http://dx.doi.org/#1}{#2}}

\bibitem[{\citenamefont{Perlmutter and Schmidt}(2003)}]{ps03}
\bibinfo{author}{\bibfnamefont{S.}~\bibnamefont{Perlmutter}} \bibnamefont{and}
  \bibinfo{author}{\bibfnamefont{B.~P.} \bibnamefont{Schmidt}},
  \bibinfo{journal}{Lect. Notes Phys.}
  \textbf{\doi{10.1007/3-540-45863-8_11}{\bibinfo{volume}{598}}},
  \doi{10.1007/3-540-45863-8_11}{\bibinfo{pages}{195}} (\bibinfo{year}{2003}),
  \eprint{astro-ph/0303428}.

\bibitem[{\citenamefont{Frieman et~al.}(2008)\citenamefont{Frieman, Turner, and
  Huterer}}]{fth08}
\bibinfo{author}{\bibfnamefont{J.}~\bibnamefont{Frieman}},
  \bibinfo{author}{\bibfnamefont{M.}~\bibnamefont{Turner}}, \bibnamefont{and}
  \bibinfo{author}{\bibfnamefont{D.}~\bibnamefont{Huterer}},
  \bibinfo{journal}{Ann. Rev. Astron. Astrophys.}
  \textbf{\doi{10.1146/annurev.astro.46.060407.145243}{\bibinfo{volume}{46}}},
  \doi{10.1146/annurev.astro.46.060407.145243}{\bibinfo{pages}{385}}
  (\bibinfo{year}{2008}), \eprint{0803.0982} [astro-ph].

\bibitem[{\citenamefont{Linder}(2008)}]{linder08}
\bibinfo{author}{\bibfnamefont{E.~V.} \bibnamefont{Linder}},
  \bibinfo{journal}{Rept. Prog. Phys.}
  \textbf{\doi{10.1088/0034-4885/71/5/056901}{\bibinfo{volume}{71}}},
  \doi{10.1088/0034-4885/71/5/056901}{\bibinfo{pages}{056901}}
  (\bibinfo{year}{2008}), \eprint{0801.2968} [astro-ph].

\bibitem[{\citenamefont{Tomita}(2000)}]{tomita00}
\bibinfo{author}{\bibfnamefont{K.}~\bibnamefont{Tomita}},
  \bibinfo{journal}{Astrophys. J.} \textbf{\bibinfo{volume}{529}},
  \bibinfo{pages}{38} (\bibinfo{year}{2000}), \eprint{astro-ph/9906027}.

\bibitem[{\citenamefont{Goodwin et~al.}(1999)\citenamefont{Goodwin, Thomas,
  Barber, Gribbin, and Onuora}}]{gtbgo99}
\bibinfo{author}{\bibfnamefont{S.~P.} \bibnamefont{Goodwin}},
  \bibinfo{author}{\bibfnamefont{P.~A.} \bibnamefont{Thomas}},
  \bibinfo{author}{\bibfnamefont{A.~J.} \bibnamefont{Barber}},
  \bibinfo{author}{\bibfnamefont{J.}~\bibnamefont{Gribbin}}, \bibnamefont{and}
  \bibinfo{author}{\bibfnamefont{L.~I.} \bibnamefont{Onuora}}
  (\bibinfo{year}{1999}), \eprint{astro-ph/9906187}.

\bibitem[{\citenamefont{Celerier}(2000)}]{celerier99}
\bibinfo{author}{\bibfnamefont{M.-N.} \bibnamefont{Celerier}},
  \bibinfo{journal}{Astron. Astrophys.} \textbf{\bibinfo{volume}{353}},
  \bibinfo{pages}{63} (\bibinfo{year}{2000}), \eprint{astro-ph/9907206}.

\bibitem[{\citenamefont{{Moffat} and {Tatarski}}(1995)}]{moffat95}
\bibinfo{author}{\bibfnamefont{J.~W.} \bibnamefont{{Moffat}}} \bibnamefont{and}
  \bibinfo{author}{\bibfnamefont{D.~C.} \bibnamefont{{Tatarski}}},
  \bibinfo{journal}{Astrophys. J.}
  \textbf{\doi{10.1086/176365}{\bibinfo{volume}{453}}},
  \doi{10.1086/176365}{\bibinfo{pages}{17}} (\bibinfo{year}{1995}),
  \eprint{astro-ph/9407036}.

\bibitem[{\citenamefont{Nakao et~al.}(1995)}]{nakao95}
\bibinfo{author}{\bibfnamefont{K.-i.} \bibnamefont{Nakao}}
  \bibnamefont{et~al.}, \bibinfo{journal}{Astrophys. J.}
  \textbf{\doi{10.1086/176416}{\bibinfo{volume}{453}}},
  \doi{10.1086/176416}{\bibinfo{pages}{541}} (\bibinfo{year}{1995}),
  \eprint{astro-ph/9502054}.

\bibitem[{\citenamefont{{Tomita}}(1996)}]{tomita96}
\bibinfo{author}{\bibfnamefont{K.}~\bibnamefont{{Tomita}}},
  \bibinfo{journal}{Astrophys. J.}
  \textbf{\doi{10.1086/177077}{\bibinfo{volume}{461}}},
  \doi{10.1086/177077}{\bibinfo{pages}{507}} (\bibinfo{year}{1996}).

\bibitem[{\citenamefont{Humphreys et~al.}(1997)\citenamefont{Humphreys,
  Maartens, and Matravers}}]{hmm97}
\bibinfo{author}{\bibfnamefont{N.~P.} \bibnamefont{Humphreys}},
  \bibinfo{author}{\bibfnamefont{R.}~\bibnamefont{Maartens}}, \bibnamefont{and}
  \bibinfo{author}{\bibfnamefont{D.~R.} \bibnamefont{Matravers}},
  \bibinfo{journal}{Astrophys. J.}
  \textbf{\doi{10.1086/303672}{\bibinfo{volume}{477}}},
  \doi{10.1086/303672}{\bibinfo{pages}{47}} (\bibinfo{year}{1997}),
  \eprint{astro-ph/9602033}.

\bibitem[{\citenamefont{Foreman et~al.}(2010)\citenamefont{Foreman, Moss,
  Zibin, and Scott}}]{fmzs10}
\bibinfo{author}{\bibfnamefont{S.}~\bibnamefont{Foreman}},
  \bibinfo{author}{\bibfnamefont{A.}~\bibnamefont{Moss}},
  \bibinfo{author}{\bibfnamefont{J.~P.} \bibnamefont{Zibin}}, \bibnamefont{and}
  \bibinfo{author}{\bibfnamefont{D.}~\bibnamefont{Scott}},
  \bibinfo{journal}{Phys. Rev.}
  \textbf{\doi{10.1103/PhysRevD.82.103532}{\bibinfo{volume}{D82}}},
  \doi{10.1103/PhysRevD.82.103532}{\bibinfo{pages}{103532}}
  (\bibinfo{year}{2010}), \eprint{1009.0273} [astro-ph.CO].

\bibitem[{\citenamefont{Marra and Notari}(2011)}]{mn11}
\bibinfo{author}{\bibfnamefont{V.}~\bibnamefont{Marra}} \bibnamefont{and}
  \bibinfo{author}{\bibfnamefont{A.}~\bibnamefont{Notari}}
  (\bibinfo{year}{2011}), \eprint{1102.1015} [astro-ph.CO].

\bibitem[{\citenamefont{Bolejko et~al.}(2011)\citenamefont{Bolejko, Celerier,
  and Krasinski}}]{bck11}
\bibinfo{author}{\bibfnamefont{K.}~\bibnamefont{Bolejko}},
  \bibinfo{author}{\bibfnamefont{M.-N.} \bibnamefont{Celerier}},
  \bibnamefont{and} \bibinfo{author}{\bibfnamefont{A.}~\bibnamefont{Krasinski}}
  (\bibinfo{year}{2011}), \eprint{1102.1449} [astro-ph.CO].

\bibitem[{\citenamefont{Yoo et~al.}(2008)\citenamefont{Yoo, Kai, and
  Nakao}}]{ykn08}
\bibinfo{author}{\bibfnamefont{C.-M.} \bibnamefont{Yoo}},
  \bibinfo{author}{\bibfnamefont{T.}~\bibnamefont{Kai}}, \bibnamefont{and}
  \bibinfo{author}{\bibfnamefont{K.-i.} \bibnamefont{Nakao}},
  \bibinfo{journal}{Prog. Theor. Phys.}
  \textbf{\doi{10.1143/PTP.120.937}{\bibinfo{volume}{120}}},
  \doi{10.1143/PTP.120.937}{\bibinfo{pages}{937}} (\bibinfo{year}{2008}),
  \eprint{0807.0932} [astro-ph].

\bibitem[{\citenamefont{Garcia-Bellido and
  Haugboelle}(2008{\natexlab{a}})}]{gbh08}
\bibinfo{author}{\bibfnamefont{J.}~\bibnamefont{Garcia-Bellido}}
  \bibnamefont{and}
  \bibinfo{author}{\bibfnamefont{T.}~\bibnamefont{Haugboelle}},
  \bibinfo{journal}{JCAP}
  \textbf{\doi{10.1088/1475-7516/2008/04/003}{\bibinfo{volume}{0804}}},
  \doi{10.1088/1475-7516/2008/04/003}{\bibinfo{pages}{003}}
  (\bibinfo{year}{2008}{\natexlab{a}}), \eprint{0802.1523} [astro-ph].

\bibitem[{\citenamefont{Zibin et~al.}(2008)\citenamefont{Zibin, Moss, and
  Scott}}]{zms08}
\bibinfo{author}{\bibfnamefont{J.~P.} \bibnamefont{Zibin}},
  \bibinfo{author}{\bibfnamefont{A.}~\bibnamefont{Moss}}, \bibnamefont{and}
  \bibinfo{author}{\bibfnamefont{D.}~\bibnamefont{Scott}},
  \bibinfo{journal}{Phys. Rev. Lett.}
  \textbf{\doi{10.1103/PhysRevLett.101.251303}{\bibinfo{volume}{101}}},
  \doi{10.1103/PhysRevLett.101.251303}{\bibinfo{pages}{251303}}
  (\bibinfo{year}{2008}), \eprint{0809.3761} [astro-ph].

\bibitem[{\citenamefont{Zibin}(2008)}]{z08}
\bibinfo{author}{\bibfnamefont{J.~P.} \bibnamefont{Zibin}},
  \bibinfo{journal}{Phys. Rev.}
  \textbf{\doi{10.1103/PhysRevD.78.043504}{\bibinfo{volume}{D78}}},
  \doi{10.1103/PhysRevD.78.043504}{\bibinfo{pages}{043504}}
  (\bibinfo{year}{2008}), \eprint{0804.1787} [astro-ph].

\bibitem[{\citenamefont{Clarkson et~al.}(2009)\citenamefont{Clarkson, Clifton,
  and February}}]{ccf09}
\bibinfo{author}{\bibfnamefont{C.}~\bibnamefont{Clarkson}},
  \bibinfo{author}{\bibfnamefont{T.}~\bibnamefont{Clifton}}, \bibnamefont{and}
  \bibinfo{author}{\bibfnamefont{S.}~\bibnamefont{February}},
  \bibinfo{journal}{JCAP}
  \textbf{\doi{10.1088/1475-7516/2009/06/025}{\bibinfo{volume}{0906}}},
  \doi{10.1088/1475-7516/2009/06/025}{\bibinfo{pages}{025}}
  (\bibinfo{year}{2009}), \eprint{0903.5040} [astro-ph.CO].

\bibitem[{\citenamefont{Moss et~al.}(2011)\citenamefont{Moss, Zibin, and
  Scott}}]{mzs10}
\bibinfo{author}{\bibfnamefont{A.}~\bibnamefont{Moss}},
  \bibinfo{author}{\bibfnamefont{J.~P.} \bibnamefont{Zibin}}, \bibnamefont{and}
  \bibinfo{author}{\bibfnamefont{D.}~\bibnamefont{Scott}},
  \bibinfo{journal}{Phys. Rev.}
  \textbf{\doi{10.1103/PhysRevD.83.103515}{\bibinfo{volume}{D83}}},
  \doi{10.1103/PhysRevD.83.103515}{\bibinfo{pages}{103515}}
  (\bibinfo{year}{2011}), \eprint{1007.3725} [astro-ph.CO].

\bibitem[{\citenamefont{Biswas et~al.}(2010)\citenamefont{Biswas, Notari, and
  Valkenburg}}]{bnv10}
\bibinfo{author}{\bibfnamefont{T.}~\bibnamefont{Biswas}},
  \bibinfo{author}{\bibfnamefont{A.}~\bibnamefont{Notari}}, \bibnamefont{and}
  \bibinfo{author}{\bibfnamefont{W.}~\bibnamefont{Valkenburg}},
  \bibinfo{journal}{JCAP}
  \textbf{\doi{10.1088/1475-7516/2010/11/030}{\bibinfo{volume}{1011}}},
  \doi{10.1088/1475-7516/2010/11/030}{\bibinfo{pages}{030}}
  (\bibinfo{year}{2010}), \eprint{1007.3065} [astro-ph.CO].

\bibitem[{\citenamefont{Marra and Paakkonen}(2010)}]{mp10}
\bibinfo{author}{\bibfnamefont{V.}~\bibnamefont{Marra}} \bibnamefont{and}
  \bibinfo{author}{\bibfnamefont{M.}~\bibnamefont{Paakkonen}},
  \bibinfo{journal}{JCAP}
  \textbf{\doi{10.1088/1475-7516/2010/12/021}{\bibinfo{volume}{1012}}},
  \doi{10.1088/1475-7516/2010/12/021}{\bibinfo{pages}{021}}
  (\bibinfo{year}{2010}), \eprint{1009.4193} [astro-ph.CO].

\bibitem[{\citenamefont{Sunyaev and Zeldovich}(1980)}]{sz80}
\bibinfo{author}{\bibfnamefont{R.~A.} \bibnamefont{Sunyaev}} \bibnamefont{and}
  \bibinfo{author}{\bibfnamefont{Y.~B.} \bibnamefont{Zeldovich}},
  \bibinfo{journal}{Mon. Not. Roy. Astron. Soc.}
  \textbf{\bibinfo{volume}{190}}, \bibinfo{pages}{413} (\bibinfo{year}{1980}).

\bibitem[{\citenamefont{Vanderveld et~al.}(2006)\citenamefont{Vanderveld,
  Flanagan, and Wasserman}}]{vfw06}
\bibinfo{author}{\bibfnamefont{R.~A.} \bibnamefont{Vanderveld}},
  \bibinfo{author}{\bibfnamefont{E.~E.} \bibnamefont{Flanagan}},
  \bibnamefont{and}
  \bibinfo{author}{\bibfnamefont{I.}~\bibnamefont{Wasserman}},
  \bibinfo{journal}{Phys. Rev.}
  \textbf{\doi{10.1103/PhysRevD.74.023506}{\bibinfo{volume}{D74}}},
  \doi{10.1103/PhysRevD.74.023506}{\bibinfo{pages}{023506}}
  (\bibinfo{year}{2006}), \eprint{astro-ph/0602476}.

\bibitem[{\citenamefont{Garcia-Bellido and
  Haugboelle}(2008{\natexlab{b}})}]{gbh08b}
\bibinfo{author}{\bibfnamefont{J.}~\bibnamefont{Garcia-Bellido}}
  \bibnamefont{and}
  \bibinfo{author}{\bibfnamefont{T.}~\bibnamefont{Haugboelle}},
  \bibinfo{journal}{JCAP}
  \textbf{\doi{10.1088/1475-7516/2008/09/016}{\bibinfo{volume}{0809}}},
  \doi{10.1088/1475-7516/2008/09/016}{\bibinfo{pages}{016}}
  (\bibinfo{year}{2008}{\natexlab{b}}), \eprint{0807.1326} [astro-ph].

\bibitem[{\citenamefont{Yoo et~al.}(2010)\citenamefont{Yoo, Nakao, and
  Sasaki}}]{yns10b}
\bibinfo{author}{\bibfnamefont{C.-M.} \bibnamefont{Yoo}},
  \bibinfo{author}{\bibfnamefont{K.-i.} \bibnamefont{Nakao}}, \bibnamefont{and}
  \bibinfo{author}{\bibfnamefont{M.}~\bibnamefont{Sasaki}},
  \bibinfo{journal}{JCAP}
  \textbf{\doi{10.1088/1475-7516/2010/10/011}{\bibinfo{volume}{1010}}},
  \doi{10.1088/1475-7516/2010/10/011}{\bibinfo{pages}{011}}
  (\bibinfo{year}{2010}), \eprint{1008.0469} [astro-ph.CO].

\bibitem[{\citenamefont{Zhang and Stebbins}(2010)}]{zs10}
\bibinfo{author}{\bibfnamefont{P.}~\bibnamefont{Zhang}} \bibnamefont{and}
  \bibinfo{author}{\bibfnamefont{A.}~\bibnamefont{Stebbins}}
  (\bibinfo{year}{2010}), \eprint{1009.3967} [astro-ph.CO].

\bibitem[{\citenamefont{{Lema{\^i}tre}}(1933)}]{lemaitre33}
\bibinfo{author}{\bibfnamefont{G.}~\bibnamefont{{Lema{\^i}tre}}},
  \bibinfo{journal}{Ann. Soc. Sci. Bruxelles} \textbf{\bibinfo{volume}{53}},
  \bibinfo{pages}{51} (\bibinfo{year}{1933}).

\bibitem[{\citenamefont{Tolman}(1934)}]{tolman34}
\bibinfo{author}{\bibfnamefont{R.~C.} \bibnamefont{Tolman}},
  \bibinfo{journal}{Proc. Nat. Acad. Sci.} \textbf{\bibinfo{volume}{20}},
  \bibinfo{pages}{169} (\bibinfo{year}{1934}).

\bibitem[{\citenamefont{Bondi}(1947)}]{bondi47}
\bibinfo{author}{\bibfnamefont{H.}~\bibnamefont{Bondi}}, \bibinfo{journal}{Mon.
  Not. Roy. Astron. Soc.} \textbf{\bibinfo{volume}{107}}, \bibinfo{pages}{410}
  (\bibinfo{year}{1947}).

\bibitem[{\citenamefont{{Kaiser}}(1984)}]{kaiser84}
\bibinfo{author}{\bibfnamefont{N.}~\bibnamefont{{Kaiser}}},
  \bibinfo{journal}{Astrophys. J.}
  \textbf{\doi{10.1086/162213}{\bibinfo{volume}{282}}},
  \doi{10.1086/162213}{\bibinfo{pages}{374}} (\bibinfo{year}{1984}).

\bibitem[{\citenamefont{{Ostriker} and {Vishniac}}(1986)}]{ov86}
\bibinfo{author}{\bibfnamefont{J.~P.} \bibnamefont{{Ostriker}}}
  \bibnamefont{and} \bibinfo{author}{\bibfnamefont{E.~T.}
  \bibnamefont{{Vishniac}}}, \bibinfo{journal}{Astrophys. J. Lett.}
  \textbf{\doi{10.1086/184704}{\bibinfo{volume}{306}}},
  \doi{10.1086/184704}{\bibinfo{pages}{L51}} (\bibinfo{year}{1986}).

\bibitem[{\citenamefont{Larson et~al.}(2011)}]{wmap7params}
\bibinfo{author}{\bibfnamefont{D.}~\bibnamefont{Larson}} \bibnamefont{et~al.},
  \bibinfo{journal}{Astrophys. J. Suppl.}
  \textbf{\doi{10.1088/0067-0049/192/2/16}{\bibinfo{volume}{192}}},
  \doi{10.1088/0067-0049/192/2/16}{\bibinfo{pages}{16}} (\bibinfo{year}{2011}),
  \eprint{1001.4635} [astro-ph.CO].

\bibitem[{\citenamefont{Eisenstein et~al.}(2005)}]{eisenstein05}
\bibinfo{author}{\bibfnamefont{D.~J.} \bibnamefont{Eisenstein}}
  \bibnamefont{et~al.} (\bibinfo{collaboration}{SDSS}),
  \bibinfo{journal}{Astrophys. J.}
  \textbf{\doi{10.1086/466512}{\bibinfo{volume}{633}}},
  \doi{10.1086/466512}{\bibinfo{pages}{560}} (\bibinfo{year}{2005}),
  \eprint{astro-ph/0501171}.

\bibitem[{\citenamefont{{Wickramasinghe} and {Ukwatta}}(2010)}]{wu10}
\bibinfo{author}{\bibfnamefont{T.}~\bibnamefont{{Wickramasinghe}}}
  \bibnamefont{and} \bibinfo{author}{\bibfnamefont{T.~N.}
  \bibnamefont{{Ukwatta}}}, \bibinfo{journal}{Mon. Not. Roy. Astron. Soc.}
  \textbf{\doi{10.1111/j.1365-2966.2010.16686.x}{\bibinfo{volume}{406}}},
  \doi{10.1111/j.1365-2966.2010.16686.x}{\bibinfo{pages}{548}}
  (\bibinfo{year}{2010}), \eprint{1003.0483} [astro-ph.CO].

\bibitem[{\citenamefont{Lewis et~al.}(2000)\citenamefont{Lewis, Challinor, and
  Lasenby}}]{camb}
\bibinfo{author}{\bibfnamefont{A.}~\bibnamefont{Lewis}},
  \bibinfo{author}{\bibfnamefont{A.}~\bibnamefont{Challinor}},
  \bibnamefont{and} \bibinfo{author}{\bibfnamefont{A.}~\bibnamefont{Lasenby}},
  \bibinfo{journal}{Astrophys. J.}
  \textbf{\doi{10.1086/309179}{\bibinfo{volume}{538}}},
  \doi{10.1086/309179}{\bibinfo{pages}{473}} (\bibinfo{year}{2000}),
  \eprint{astro-ph/9911177}.

\bibitem[{\citenamefont{Smith et~al.}(2003)}]{halofit}
\bibinfo{author}{\bibfnamefont{R.~E.} \bibnamefont{Smith}} \bibnamefont{et~al.}
  (\bibinfo{collaboration}{The Virgo Consortium}), \bibinfo{journal}{Mon. Not.
  Roy. Astron. Soc.}
  \textbf{\doi{10.1046/j.1365-8711.2003.06503.x}{\bibinfo{volume}{341}}},
  \doi{10.1046/j.1365-8711.2003.06503.x}{\bibinfo{pages}{1311}}
  (\bibinfo{year}{2003}), \eprint{astro-ph/0207664}.

\bibitem[{\citenamefont{Mustapha et~al.}(1998)\citenamefont{Mustapha, Bassett,
  Hellaby, and Ellis}}]{mbhe98}
\bibinfo{author}{\bibfnamefont{N.}~\bibnamefont{Mustapha}},
  \bibinfo{author}{\bibfnamefont{B.~A.} \bibnamefont{Bassett}},
  \bibinfo{author}{\bibfnamefont{C.}~\bibnamefont{Hellaby}}, \bibnamefont{and}
  \bibinfo{author}{\bibfnamefont{G.~F.~R.} \bibnamefont{Ellis}},
  \bibinfo{journal}{Class. Quant. Grav.}
  \textbf{\doi{10.1088/0264-9381/15/8/016}{\bibinfo{volume}{15}}},
  \doi{10.1088/0264-9381/15/8/016}{\bibinfo{pages}{2363}}
  (\bibinfo{year}{1998}), \eprint{gr-qc/9708043}.

\bibitem[{\citenamefont{{Limber}}(1953)}]{limber53}
\bibinfo{author}{\bibfnamefont{D.~N.} \bibnamefont{{Limber}}},
  \bibinfo{journal}{Astrophys. J.}
  \textbf{\doi{10.1086/145672}{\bibinfo{volume}{117}}},
  \doi{10.1086/145672}{\bibinfo{pages}{134}} (\bibinfo{year}{1953}).

\bibitem[{\citenamefont{Afshordi et~al.}(2004)\citenamefont{Afshordi, Loh, and
  Strauss}}]{als04}
\bibinfo{author}{\bibfnamefont{N.}~\bibnamefont{Afshordi}},
  \bibinfo{author}{\bibfnamefont{Y.-S.} \bibnamefont{Loh}}, \bibnamefont{and}
  \bibinfo{author}{\bibfnamefont{M.~A.} \bibnamefont{Strauss}},
  \bibinfo{journal}{Phys. Rev.}
  \textbf{\doi{10.1103/PhysRevD.69.083524}{\bibinfo{volume}{D69}}},
  \doi{10.1103/PhysRevD.69.083524}{\bibinfo{pages}{083524}}
  (\bibinfo{year}{2004}), \eprint{astro-ph/0308260}.

\bibitem[{\citenamefont{Amanullah et~al.}(2010)}]{Amanullah:2010vv}
\bibinfo{author}{\bibfnamefont{R.}~\bibnamefont{Amanullah}}
  \bibnamefont{et~al.}, \bibinfo{journal}{Astrophys. J.}
  \textbf{\doi{10.1088/0004-637X/716/1/712}{\bibinfo{volume}{716}}},
  \doi{10.1088/0004-637X/716/1/712}{\bibinfo{pages}{712}}
  (\bibinfo{year}{2010}), \eprint{1004.1711} [astro-ph.CO].

\bibitem[{\citenamefont{Freedman et~al.}(2001)}]{Freedman:2000cf}
\bibinfo{author}{\bibfnamefont{W.~L.} \bibnamefont{Freedman}}
  \bibnamefont{et~al.} (\bibinfo{collaboration}{HST}),
  \bibinfo{journal}{Astrophys. J.}
  \textbf{\doi{10.1086/320638}{\bibinfo{volume}{553}}},
  \doi{10.1086/320638}{\bibinfo{pages}{47}} (\bibinfo{year}{2001}),
  \eprint{astro-ph/0012376}.

\bibitem[{\citenamefont{Riess et~al.}(2009)}]{riessetal09}
\bibinfo{author}{\bibfnamefont{A.~G.} \bibnamefont{Riess}}
  \bibnamefont{et~al.}, \bibinfo{journal}{Astrophys. J.}
  \textbf{\doi{10.1088/0004-637X/699/1/539}{\bibinfo{volume}{699}}},
  \doi{10.1088/0004-637X/699/1/539}{\bibinfo{pages}{539}}
  (\bibinfo{year}{2009}), \eprint{0905.0695} [astro-ph.CO].

\bibitem[{\citenamefont{{Riess} et~al.}(2011)}]{riessetal11}
\bibinfo{author}{\bibfnamefont{A.~G.} \bibnamefont{{Riess}}}
  \bibnamefont{et~al.}, \bibinfo{journal}{Astrophys. J.}
  \textbf{\doi{10.1088/0004-637X/730/2/119}{\bibinfo{volume}{730}}},
  \doi{10.1088/0004-637X/730/2/119}{\bibinfo{pages}{119}}
  (\bibinfo{year}{2011}), \eprint{1103.2976} [astro-ph.CO].

\bibitem[{\citenamefont{Nadathur and Sarkar}(2011)}]{ns11}
\bibinfo{author}{\bibfnamefont{S.}~\bibnamefont{Nadathur}} \bibnamefont{and}
  \bibinfo{author}{\bibfnamefont{S.}~\bibnamefont{Sarkar}},
  \bibinfo{journal}{Phys. Rev.}
  \textbf{\doi{10.1103/PhysRevD.83.063506}{\bibinfo{volume}{D83}}},
  \doi{10.1103/PhysRevD.83.063506}{\bibinfo{pages}{063506}}
  (\bibinfo{year}{2011}), \eprint{1012.3460} [astro-ph.CO].

\bibitem[{\citenamefont{Lewis and Bridle}(2002)}]{cosmomc}
\bibinfo{author}{\bibfnamefont{A.}~\bibnamefont{Lewis}} \bibnamefont{and}
  \bibinfo{author}{\bibfnamefont{S.}~\bibnamefont{Bridle}},
  \bibinfo{journal}{Phys. Rev.}
  \textbf{\doi{10.1103/PhysRevD.66.103511}{\bibinfo{volume}{D66}}},
  \doi{10.1103/PhysRevD.66.103511}{\bibinfo{pages}{103511}}
  (\bibinfo{year}{2002}), \eprint{astro-ph/0205436}.

\bibitem[{\citenamefont{Carlberg et~al.}(1997)}]{carlbergetal97}
\bibinfo{author}{\bibfnamefont{R.~G.} \bibnamefont{Carlberg}}
  \bibnamefont{et~al.} (\bibinfo{year}{1997}), \eprint{astro-ph/9711272}.

\bibitem[{\citenamefont{Fukugita et~al.}(1998)\citenamefont{Fukugita, Hogan,
  and Peebles}}]{fhp98}
\bibinfo{author}{\bibfnamefont{M.}~\bibnamefont{Fukugita}},
  \bibinfo{author}{\bibfnamefont{C.~J.} \bibnamefont{Hogan}}, \bibnamefont{and}
  \bibinfo{author}{\bibfnamefont{P.~J.~E.} \bibnamefont{Peebles}},
  \bibinfo{journal}{Astrophys. J.}
  \textbf{\doi{10.1086/306025}{\bibinfo{volume}{503}}},
  \doi{10.1086/306025}{\bibinfo{pages}{518}} (\bibinfo{year}{1998}),
  \eprint{astro-ph/9712020}.

\bibitem[{\citenamefont{Reichardt et~al.}(2009)}]{Reichardt:2008ay}
\bibinfo{author}{\bibfnamefont{C.~L.} \bibnamefont{Reichardt}}
  \bibnamefont{et~al.}, \bibinfo{journal}{Astrophys. J.}
  \textbf{\doi{10.1088/0004-637X/694/2/1200}{\bibinfo{volume}{694}}},
  \doi{10.1088/0004-637X/694/2/1200}{\bibinfo{pages}{1200}}
  (\bibinfo{year}{2009}), \eprint{0801.1491} [astro-ph].

\bibitem[{\citenamefont{Hall et~al.}(2010)}]{Hall:2009rv}
\bibinfo{author}{\bibfnamefont{N.~R.} \bibnamefont{Hall}} \bibnamefont{et~al.},
  \bibinfo{journal}{Astrophys. J.}
  \textbf{\doi{10.1088/0004-637X/718/2/632}{\bibinfo{volume}{718}}},
  \doi{10.1088/0004-637X/718/2/632}{\bibinfo{pages}{632}}
  (\bibinfo{year}{2010}), \eprint{0912.4315} [astro-ph.CO].

\bibitem[{\citenamefont{Shirokoff et~al.}(2011)}]{shirokoffetal10}
\bibinfo{author}{\bibfnamefont{E.}~\bibnamefont{Shirokoff}}
  \bibnamefont{et~al.}, \bibinfo{journal}{Astrophys. J.}
  \textbf{\doi{10.1088/0004-637X/736/1/61}{\bibinfo{volume}{736}}},
  \doi{10.1088/0004-637X/736/1/61}{\bibinfo{pages}{61}} (\bibinfo{year}{2011}),
  \eprint{1012.4788} [astro-ph.CO].

\bibitem[{\citenamefont{Das et~al.}(2011)}]{Das:2010ga}
\bibinfo{author}{\bibfnamefont{S.}~\bibnamefont{Das}} \bibnamefont{et~al.},
  \bibinfo{journal}{Astrophys. J.}
  \textbf{\doi{10.1088/0004-637X/729/1/62}{\bibinfo{volume}{729}}},
  \doi{10.1088/0004-637X/729/1/62}{\bibinfo{pages}{62}} (\bibinfo{year}{2011}),
  \eprint{1009.0847} [astro-ph.CO].

\bibitem[{\citenamefont{Alnes and Amarzguioui}(2006)}]{aa06}
\bibinfo{author}{\bibfnamefont{H.}~\bibnamefont{Alnes}} \bibnamefont{and}
  \bibinfo{author}{\bibfnamefont{M.}~\bibnamefont{Amarzguioui}},
  \bibinfo{journal}{Phys. Rev.}
  \textbf{\doi{10.1103/PhysRevD.74.103520}{\bibinfo{volume}{D74}}},
  \doi{10.1103/PhysRevD.74.103520}{\bibinfo{pages}{103520}}
  (\bibinfo{year}{2006}), \eprint{astro-ph/0607334}.

\bibitem[{\citenamefont{Dore et~al.}(2004)\citenamefont{Dore, Hennawi, and
  Spergel}}]{dhs04}
\bibinfo{author}{\bibfnamefont{O.}~\bibnamefont{Dore}},
  \bibinfo{author}{\bibfnamefont{J.~F.} \bibnamefont{Hennawi}},
  \bibnamefont{and} \bibinfo{author}{\bibfnamefont{D.~N.}
  \bibnamefont{Spergel}}, \bibinfo{journal}{Astrophys. J.}
  \textbf{\doi{10.1086/382946}{\bibinfo{volume}{606}}},
  \doi{10.1086/382946}{\bibinfo{pages}{46}} (\bibinfo{year}{2004}),
  \eprint{astro-ph/0309337}.

\bibitem[{\citenamefont{Fu et~al.}(2008)}]{fuetal08}
\bibinfo{author}{\bibfnamefont{L.}~\bibnamefont{Fu}} \bibnamefont{et~al.},
  \bibinfo{journal}{Astron. Astrophys.}
  \textbf{\doi{10.1051/0004-6361:20078522}{\bibinfo{volume}{479}}},
  \doi{10.1051/0004-6361:20078522}{\bibinfo{pages}{9}} (\bibinfo{year}{2008}),
  \eprint{0712.0884} [astro-ph].

\bibitem[{\citenamefont{Clarkson and Regis}(2011)}]{cr10}
\bibinfo{author}{\bibfnamefont{C.}~\bibnamefont{Clarkson}} \bibnamefont{and}
  \bibinfo{author}{\bibfnamefont{M.}~\bibnamefont{Regis}},
  \bibinfo{journal}{JCAP} \textbf{\bibinfo{volume}{1102}}, \bibinfo{pages}{013}
  (\bibinfo{year}{2011}), \eprint{1007.3443} [astro-ph.CO].

\bibitem[{\citenamefont{Jing et~al.}(2006)\citenamefont{Jing, Zhang, Lin, Gao,
  and Springel}}]{jzlgs06}
\bibinfo{author}{\bibfnamefont{Y.~P.} \bibnamefont{Jing}},
  \bibinfo{author}{\bibfnamefont{P.}~\bibnamefont{Zhang}},
  \bibinfo{author}{\bibfnamefont{W.~P.} \bibnamefont{Lin}},
  \bibinfo{author}{\bibfnamefont{L.}~\bibnamefont{Gao}}, \bibnamefont{and}
  \bibinfo{author}{\bibfnamefont{V.}~\bibnamefont{Springel}},
  \bibinfo{journal}{Astrophys. J.}
  \textbf{\doi{10.1086/503547}{\bibinfo{volume}{640}}},
  \doi{10.1086/503547}{\bibinfo{pages}{L119}} (\bibinfo{year}{2006}),
  \eprint{astro-ph/0512426}.

\bibitem[{\citenamefont{DeDeo et~al.}(2005)\citenamefont{DeDeo, Spergel, and
  Trac}}]{dst05}
\bibinfo{author}{\bibfnamefont{S.}~\bibnamefont{DeDeo}},
  \bibinfo{author}{\bibfnamefont{D.~N.} \bibnamefont{Spergel}},
  \bibnamefont{and} \bibinfo{author}{\bibfnamefont{H.}~\bibnamefont{Trac}}
  (\bibinfo{year}{2005}), \eprint{astro-ph/0511060}.

\bibitem[{\citenamefont{Goodman}(1995)}]{goodman95}
\bibinfo{author}{\bibfnamefont{J.}~\bibnamefont{Goodman}},
  \bibinfo{journal}{Phys. Rev.}
  \textbf{\doi{10.1103/PhysRevD.52.1821}{\bibinfo{volume}{D52}}},
  \doi{10.1103/PhysRevD.52.1821}{\bibinfo{pages}{1821}} (\bibinfo{year}{1995}),
  \eprint{astro-ph/9506068}.

\bibitem[{\citenamefont{Caldwell and Stebbins}(2008)}]{cs08}
\bibinfo{author}{\bibfnamefont{R.~R.} \bibnamefont{Caldwell}} \bibnamefont{and}
  \bibinfo{author}{\bibfnamefont{A.}~\bibnamefont{Stebbins}},
  \bibinfo{journal}{Phys. Rev. Lett.}
  \textbf{\doi{10.1103/PhysRevLett.100.191302}{\bibinfo{volume}{100}}},
  \doi{10.1103/PhysRevLett.100.191302}{\bibinfo{pages}{191302}}
  (\bibinfo{year}{2008}), \eprint{0711.3459} [astro-ph].

\bibitem[{\citenamefont{Fixsen et~al.}(1996)}]{fixsenetal96}
\bibinfo{author}{\bibfnamefont{D.~J.} \bibnamefont{Fixsen}}
  \bibnamefont{et~al.}, \bibinfo{journal}{Astrophys. J.}
  \textbf{\doi{10.1086/178173}{\bibinfo{volume}{473}}},
  \doi{10.1086/178173}{\bibinfo{pages}{576}} (\bibinfo{year}{1996}),
  \eprint{astro-ph/9605054}.

\bibitem[{\citenamefont{{Silk}}(1977)}]{silk77}
\bibinfo{author}{\bibfnamefont{J.}~\bibnamefont{{Silk}}},
  \bibinfo{journal}{Astron. Astrophys.} \textbf{\bibinfo{volume}{59}},
  \bibinfo{pages}{53} (\bibinfo{year}{1977}).

\end{thebibliography}

\end{document}